\documentclass[draft]{article}
\def\today{13.10.04} 
\usepackage{amsmath,amsfonts,amsthm,amssymb,amscd}

\newcommand{\beq}{\begin{equation}}
\newcommand{\ee}{\end{equation}}

\theoremstyle{plain} \newtheorem{theorem}{Theorem}[section]
\newtheorem{lemma}[theorem]{Lemma}
\newtheorem{proposition}[theorem]{Proposition}
\newtheorem{corollary}[theorem]{Corollary} \theoremstyle{definition}
\newtheorem{definition}[theorem]{Definition} \theoremstyle{remark}
\newtheorem{remark}[theorem]{Remark}
\newtheorem{example}[theorem]{Example}
 
\newcommand{\R}{{\mathbb R}} \newcommand{\U}{{\mathcal U}}
\newcommand{\Hc}{{\mathcal H}} 
 
\newcommand{\Z}{{\mathbb Z}}
\newcommand{\Nn}{{\mathbb N}}
\newcommand{\In}{{\mathbb Z}}
\newcommand{\be}{{\bf e}}
\newcommand{\Td}{{\T^d}}
\newcommand{\Zb}{{\bar \In}}

\newcommand{\Tr}{{\mathcal T}}

\newcommand{\Ph}{{\mathcal P}}
\newcommand{\B}{{\mathcal B}}
\newcommand{\I}{{\mathcal I}}
\newcommand{\resto}{{\mathcal
R}}
\def\im{{\rm i}}
\newcommand{\N}{{\mathcal N}} \newcommand{\Nc}{{\mathcal N}} 
\newcommand{\Sc}{{\mathcal S}}
\newcommand{\Scr}{{\mathcal S}_{r}}
\newcommand{\V}{{\mathcal V}}\newcommand{\J}{{\mathcal J}}
\newcommand{\C}{\mathbb{C}}
\newcommand{\T}{\mathbb{T}}

\newcommand{\va}[1]{|#1|}
  \newcommand{\vak}{|k|}
  \newcommand{\vaj}{|j|}
  \newcommand{\val}{|l|}
  
  \newcommand{\Va}[1]{\left|#1\right|}
\newcommand{\Ns}[1]{\|#1\|_{s}}

\newcommand{\Ze}{{\mathcal Z}}
\def\norma#1{\left\| #1\right\|}

\def\homega{H_0}
\def\phiv{\varphi}
\def\snorma#1{\left\langle\left|#1  \right|\right\rangle }
\def\mod#1{\left\lfloor#1\right\rceil}
\def\poisson#1#2{\left\{#1;#2\right\}}
\def\newton#1#2{\left(\begin{matrix}#1\cr#2\end{matrix}  \right)}
\def\zk#1{\Ze^{(#1)}}
\def\tk{\Tr^{(r)}} 
\def\fk#1{f^{(#1)}}
\def\sleq{\leq\kern-6pt \cdot\null\hskip4pt}
\def\ji{j_i}
\def\bji#1{j_{#1}}\def\bii#1{i_{#1}}
\def\der#1#2{\frac{d^{#1}\omega_{#2}}{dm^{#1}}}

\numberwithin{equation}{section}

\begin{document}

\author{D.~Bambusi, B. Gr\'ebert} \title{BIRKHOFF NORMAL FORM FOR PDEs
WITH TAME MODULUS}

\date{\today}
\maketitle

\begin{abstract}
We prove an abstract Birkhoff normal form theorem for Hamiltonian
Partial Differential Equations. The theorem applies to semilinear
equations with nonlinearity satisfying a property that we call of Tame
Modulus. Such a property is related to the classical tame inequality
by Moser. In the nonresonant case we deduce that any small amplitude
solution remains very close to a torus for very long times.  We also
develop a general scheme to apply the abstract theory to PDEs in one
space dimensions and we use it to study some concrete equations
(NLW,NLS) with different boundary conditions. An application to a
nonlinear Schr\"odinger equation on the $d$-dimensional torus is also
given.  In all cases we deduce bounds on the growth of high Sobolev
norms. In particular we get lower bounds on the existence time of
solutions.
\end{abstract}

\tableofcontents
\newpage
\setcounter{section}{0}

\section{Introduction}

During the last fifteen years remarkable results have been obtained in
perturbation theory of integrable partial differential equation. In
particular the existence of quasiperiodic solutions has been proved
through suitable extensions of KAM theory (see
\cite{K87,K1,K2,W90,Cr00,KaP,CW92, Bou03}).  However very little is
known on the behavior of the solutions lying outside KAM tori. In the
finite dimensional case a description of such solutions is provided by
Nekhoroshev's theorem whose extension to PDEs is at present a
completely open problem. In the particular case of a
neighbourhood of an elliptic equilibrium point, Birkhoff normal form
theorem provides a quite precise description of the dynamics. In the
present paper we give an extension of Birkhoff normal form theorem to
the infinite dimensional case and we apply it to some semilinear PDEs
in one or more space dimensions.

\smallskip

To start with consider a finite dimensional Hamiltonian system $H=
H_{0}+P$ with quadratic part
$$
H_{0}= \sum_{j=1}^n \omega_{j} \frac{p_{j}^{2}+q_{j}^{2}}{2}\, \quad
\omega_j\in\R 
$$ and $P$ a smooth function having a zero of order at least three at
the origin.  The classical Birkhoff normal form theorem states that,
for each $r\geq 1$, there exits a real analytic symplectic
transformation $\Tr_r$ such that
\begin{equation} \label{I.1}
H\circ \Tr_r = H_{0} +\Ze+R_{r},
\end{equation}
where the remainder $R_{r}$ has a zero of order $r+3$ and $\Ze$ is a
polynomial of degree $r+2$. Provided the frequencies are nonresonant,
$\Ze$ depends on the actions $I_{j} =(p_{j}^{2}+q_{j}^{2})/2$ only,
and one says that the Hamiltonian (\ref{I.1}) is in integrable
Birkhoff normal form up to order $r+2$.  As a dynamical consequence,
solutions with initial data of size $\varepsilon\ll 1$ remain at
distance $2\varepsilon$ from the origin for times of order
$\varepsilon^{-r}$, and the actions remain almost constant during the
same lapse of time. Moreover any solution remains $\epsilon^{r_1}$
close to a torus of maximal dimension up to times of order
$\epsilon^{-r_2}$, with $r_1+r_2=r+1$.

The proof of such a standard theorem is obtained by applying a
sequence of canonical transformations that eliminate order by order
the non normalized part of the nonlinearity. Remark that, since the
space of homogeneous polynomials of any finite order is finite
dimensional, only a finite number of monomials have to be removed. Of
course this is no more true in the infinite dimensional case.
\smallskip

To generalize Birkhoff normal form theory to infinite dimensional
systems, the main difficulty consists in finding a nonresonance
property that is satisfied in quite general situations and that allows
to remove from the nonlinearity all the relevant non-normalized
monomials. This problem has been solved by the first author in a
particular case, namely the nonlinear wave equation (see
\cite{Bam03}), by remarking that most of the monomials are ``not
relevant'', since their vector field is already small. Moreover all
the remaining monomials can be eliminated using a suitable
nonresonance condition.

In the present paper we generalize the procedure of \cite{Bam03} to
obtain an abstract result that can be applied to a wide class of
PDEs. The idea is again that a large part of the nonlinearity is `not
relevant', but we show here that this is due to the so called tame
inequality, namely
\begin{equation}\label{I.2}
\norma{uv}_{H^s}\leq C_s\left( \norma u_{H^s}\norma v_{H^1}+ 
\norma v_{H^s}\norma u_{H^1}\right)\ ,
\end{equation}   
on which we base our construction.  The point is that, if
$u=\sum_{\vaj \geq N} u_{j}e^{\im jx}$, then, by (\ref{I.2}) one has
$$
\norma{u^2}_{H^s}\leq \frac{\norma{u}_{H^s}^2}{N^{s-1}}
$$ 
which is small provided $N$ and/or $s$ are large. This remark allows
to forget a large part of the nonlinearity. However
in order to exploit (\ref{I.2}) one
has to show that the tame property is not lost along the iterative
normalization procedure. We prove that this is the case for a class of
nonlinearities that we call of {\it tame modulus} (see definition
\ref{TM}).

In our abstract theorem we show that, if the nonlinearity has tame
modulus, then it is always possible to put $H$ in the form
\eqref{I.1}, where the remainder term $R_{r}$ is of ``order" $r+5/2$
and $\Ze$ is a polynomial of degree $r+2$ containing only monomials
which are ``almost resonant" (see definition \ref{norfor}). If the
frequencies fulfill a nonresonance condition similar to the second
Melnikov condition, then $H$ can be put in integrable Birkhoff normal
form, i.e. $\Ze$ depends on the actions only. We will also show how to
deduce some informations on the dynamics when some resonances or
almost resonances are present (see sections \ref{periodic}, \ref{dnls}
below).

\smallskip

When applying this abstract Birkhoff normal form result to PDEs, one
meets two difficulties: (1) to verify the tame modulus condition;
(2) to study the structure of $\Ze$ in order to extract dynamical
informations (e.g. by showing that it depends on the actions only).

Concerning (1), in the context of PDEs with Dirichlet or periodic
boundary conditions, we obtain a simple condition ensuring tame
modulus. Thus we obtain that typically all local functions and also
convolution type nonlinearities have the tame modulus property.

The structure of $\Ze$ depends on the resonance relations among the
frequencies. Their study is quite difficult since, at variance
with respect to the case of KAM theory for PDEs, infinitely many
frequencies are actually involved. Here we prove that in the case of
NLW and NLS in one space dimensions with Dirichlet boundary conditions
$\Ze$ is typically integrable. Thus all the small initial data give
rise to solutions that remain
small and close to a torus for times of order $\varepsilon^{-r}$, where
$\varepsilon$ measures the size of the initial datum and $r$ is an
arbitrary integer (see corollary
\ref{3.9} for precise statements). 

In the case of periodic boundary conditions the nonresonance condition
is typically violated. This is due to the fact that the periodic
eigenvalues of a Sturm Liouville operator are asymptotically double.
Nevertheless we prove that, roughly speaking, the energy transfers are
allowed only between pairs of modes corresponding to almost double
eigenvalues (see theorem \ref{thm.nlwper} for a precise statement in
the case of periodic NLW equation). In particular we deduce an
estimate of higher Sobolev norms as in the Dirichlet case. We also
study a system of coupled NLS's with indefinite energy, getting the
same result as in the case of the wave equation with periodic boundary
conditions. In particular we thus obtain long time existence of
solutions somehow in the spirit of the almost global existence of
\cite{Kla83}.

Finally we apply our result to NLS in higher dimension. However,
in order to be able to control the frequencies, we have to assume that
the linear part is very simple from a spectral point of view. Namely
the linear operator we consider reads $-\Delta u + V * u$, involving a
convolution operator instead of the more natural product operator. In
this context we again obtain long time existence and bounds on the
energy exchanges among the modes (see theorem \ref{thm.nlsd}).

\smallskip

Previous results related to the present one were obtained by Bourgain
\cite{Bo00} who showed that, in perturbations of the integrable NLS,
one has that most small amplitude solutions remain close to tori for
long times. Such a result is proved by exploiting the idea that most
monomials in the nonlinearity are already small (as in the present
paper and in \cite{Bam03}). However the technique of \cite{Bo00} seems
to be quite strongly related to the particular problem dealt with in
that paper.

Other Birkhoff normal form results were presented in
\cite{Bam03a,Bam04}. Those results apply to much more general
equations (quasilinear equations in higher space dimensions) but the
kind of normal form obtained in those papers only allows to describe
the dynamics up times of order $\epsilon^{-1}$. This is due to the
fact that in the normal form of \cite{Bam03a,Bam04} the remainder
$R_r$ has a vector field which is unbounded and greatly reduces
regularity. On the contrary in the present paper the result obtained
allows to control  the dynamics for much longer times.

We point out that the study of the structure of the
resonances is based on an accurate study of the eigenvalues of
Sturm--Liouville operators. A typical difficulty met here is related to
the fact that the eigenvalues of Sturm--Liouville operators have very
rigid asymptotic properties. In order to prove our nonresonance type
conditions we use ideas from degenerate KAM theory (as in
\cite{Bam03a}) and ideas from the paper \cite{Bo96}.

\vskip20pt\noindent {\it Acknowledgements} DB was supported by the
MIUR project `Sistemi dinamici di dimensione infinita con applicazioni
ai fondamenti dinamici della meccanica statistica e alla dinamica
dell'interazione radiazione materia'. We would like to thank Thomas
Kappeler and Livio Pizzocchero for some useful discussions.

\section{Statement of the abstract result}
\label{mainres}

We will study a Hamiltonian system of the form 
\begin{eqnarray}
\label{hampq}
H(p,q):=H_0(p,q)+P(p,q)\ ,
\\
\label{h0pq}
H_0:=
\sum_{l\geq 1} \omega_l \frac{(p^{2}_{l}+q^{2}_{l})}{2} ,
\end{eqnarray}
 where the real numbers $\omega_l$ play the role of
  frequencies and $P$ has a zero of order at least three at the
  origin.  The formal Hamiltonian vector field of the system is
  $X_{H}:=(-\frac{\partial H}{\partial q_k}, \frac{\partial
  H}{\partial p_k})$.  Define the Hilbert space $\ell^{2}_s(\R)$ of
  the sequences $x\equiv \{x_l\}_{l\geq1}$ with $x_l\in\R$ such that
\begin{equation}
\label{ells}
\norma{x}_{s}^2:=\sum_{l\geq1}l^{2s}|x_l|^2<\infty
\end{equation}
and the scale of phase spaces $\Ph_s(\R):=\ell^2_s(\R)\oplus
\ell^2_s(\R)\ni(p,q)$.  We assume that $P$ is of class $C^{\infty}$
from $\Ph_s(\R)$ into $\R$ for any $s$ large enough.  We will
denote by   $z\equiv (z_l)_{l\in\Zb}$, $\Zb:=\Z-\{0\}$ the
set of all the variables, where 
$$
z_{-l}:=p_l\ ,\quad z_l:=q_l\quad l\geq 1\ .
$$
and by $B_{\R,s}(R)$ the open ball centered
at the origin and of radius $R$ in $\Ph_s(\R)$. Often we will simply
write
$$
\Ph_s \equiv \Ph_s(\R)\ ,\quad B_{s}(R) \equiv B_{\R,s}(R) \ .
$$

\begin{remark}
\label{indices}
In section \ref{appli}, depending on the concrete example we will
consider, the index $l$ will run in $\Nn$, $\Z$ or even $\Z^d$. In
this abstract section we will consider indexes in $\Nn$. Clearly one
can always reduce to this case relabelling the indexes.
\end{remark}

\subsection{Tame maps}
\label{Tmf}
Let $f:\Ph_s\to\R$ be a homogeneous polynomial of degree $r$; 
we recall that $f$ is continuous and also analytic if and only if it is
bounded, namely if there exists $C$ such that
$$
|f(z)|\leq C\norma{z}_{s}^r\ ,\quad \forall z\in\Ph_s\ .
$$ To the polynomial $f$ it is naturally associated a symmetric
$r$-linear form $\tilde f$ such that
\begin{equation}
\label{r-lin}
\tilde f(z,...,z)=f(z)\ .
\end{equation}
More explicitly, write
\begin{gather}\label{fj}
f(z)=\sum_{|j|=r}f_jz^j\ ,\quad
j=(...,j_{-l},...,j_{-1},j_1,...,j_{l},...)  \ ,
\\
\label{fj2}
z^j:=...z_{-l}^{j_{-l}}...z_{-1}^{j_{-1}} z_1^{j_1}...z_{l}^{j_l}...\
,\quad |j|:=\sum_l|j_l|\ ,
\end{gather}
then
\begin{equation}
\label{tilde}
\tilde
f(z^{(1)},...,z^{(r)})=\sum_{|j|=r}f_{j_1,...,j_r}
z^{(1)}_{j_1}... z^{(r)}_{j_r} 
\end{equation}
The multilinear form $\tilde f$ is bounded i.e.
$$
|\tilde f(z^{(1)},...,z^{(r)})|\leq
 C\norma{z^{(1)}}_{s}...\norma{z^{(r)}}_{s} 
$$
(and analytic)
if and only if $f$ is bounded.

Given a polynomial vector field $X:\Ph_s\to\Ph_s$
homogeneous of degree $r$ we write it as
$$
X(z)=\sum_{l\in\Zb}X_l(z)\be_l
$$ where $\be_l\in\Ph_s$ is the vector with all components equal to
zero but the $l-$th one which is equal to 1. Thus $X_l(z)$ is a real
valued homogeneous polynomial of degree $r$.  Consider the $r$-linear
symmetric form $\tilde X_l$ and define $\tilde X:=\sum_l\tilde
X_l\be_l$, so that
\begin{equation}
\label{r-linX}
\tilde X(z,...,z)=X(z)\ .
\end{equation}
Again $X:\Ph_s\to\Ph_s$ is analytic if and only if it is bounded.
Then the same is true for $\tilde X$.

\begin{definition}
\label{tames}
Let $X:\Ph_s\to\Ph_s$ be a homogeneous polynomial of degree $r$; let
$s\geq 1$, then we
say that $X$ is {\it an $s$--tame map} if there exists a constant
$C_s$ such that
\begin{eqnarray}
\nonumber \norma{\tilde X(z^{(1)},...,z^{(r)}) }_s &\leq& C_s
\sum_{l=1}^{r}
\norma{z^{(1)}}_{1}....\norma{z^{(l-1)}}_{1} \norma{z^{(l)}}_{s}
\norma{z^{(l+1)}}_{1}...\norma{z^{(r)}}_{1} \\ && \forall
z^{(1)},...,z^{(r)}\in\Ph_s
\label{tamemap}
\end{eqnarray}
If a map is $s$--tame for any $s\geq 1$ then it will be
said to be tame.
\qed
\end{definition}

\begin{example}
\label{ex.1} 
Given a sequence $z\equiv (z_l)_{l\in \Z}$ consider the
map
\begin{equation}
\label{t.4}
(z_l)\mapsto (w_k)\ ,\quad w_k:=\sum_{l}z_{k-l}z_l
\end{equation}
this is a tame $s$ map for any $s\geq 1$. To see this define the
map\footnote{In fact, for $z\in \Ph_s$ the index $l$ runs in
$\Z-\{0\}$, but one can reorder the indexes so that they run in $\Z$.}
\begin{equation}
\label{iso1}
\Ph_s\ni\left\{z_l\right\}\mapsto
u(x):=\sum_{l}{z_l} e^{\im lx}\in H^s(\T)
\end{equation}
where the Sobolev space $H^{s} (\T)$ is the space of periodic
functions of period $2\pi$ having $s$ weak derivatives of class
$L^2$. Then the map (\ref{t.4}) is transformed into the map
$$
u\mapsto u^2
$$ 
The corresponding bilinear
form is $(u,v)\mapsto uv$ and, by Moser inequality, one has 
\begin{equation}
\label{mos}
\norma{uv}_{H^s}\leq C_s\left( \norma u_{H^s}\norma v_{H^1}+ 
\norma v_{H^s}\norma u_{H^1}\right)
\end{equation}
\qed
\end{example}

\begin{example}
\label{linear}
According to definition \ref{tames}, any bounded linear map
$A:\Ph_s\to\Ph_s$ is $s$-tame.\qed
\end{example}

\subsection{The modulus of a map}
\label{modulus}

\begin{definition} 
\label{modulo}
Let $f$ be a homogeneous polynomial of degree $r$. Following Nikolenko
\cite{Nik86}  we define its {\it modulus}
$\mod f$ by
\begin{equation}
\label{modu1}
\mod f(z):=\sum_{|j|=r}|f_j|z^j
\end{equation}
where $f_j$ is defined by (\ref{fj}). We remark that in general the
modulus of a bounded analytic polynomial can be an unbounded densely
defined polynomial.

Analogously the modulus of a vector field $X$ is defined by 
\begin{equation}
\label{mod2}
\mod X(z):=\sum_{l\in\Zb}\mod{X_l}(z)\be_l\ .
\end{equation}
with $X_l$ the $l$-th component of $X$.\qed
\end{definition}

\begin{definition}
\label{TM}
A polynomial vector field $X$ is said to have {\it $s$--tame modulus}
if its modulus $\mod X$ is an $s$--tame map. The set of polynomial
functions $f$ whose Hamiltonian vector field has $s$--tame modulus
will be denoted by $T^s_M$, and we will write $f\in T^s_M$. If $f\in
T^s_M$ for any $s> 1$ we will write $f\in T_M$ and say that it has
tame modulus.  \qed
\end{definition}

\begin{remark}
\label{base1}
The property of having tame modulus depends on the coordinate
system.  \qed
\end{remark}

\begin{remark}
\label{m.1}
Consider a Hamiltonian function $f$ then it is easy to see that its
vector field $X_f$ has tame modulus if and only if the Hamiltonian
vector field of $\mod f$ is tame. \qed
\end{remark}

\begin{example}
\label{ex.2}
Consider again the map of example \ref{ex.1}: In this case the map
$X_k(z)$ is simply given by (\ref{t.4}) and can be written in the
form 
$$
X_k(z)=\sum_{j_1,j_2}\delta_{j_2,k-j_1}z_{j_1}z_{j_2}
$$
so that in this case $\mod X=X$, and therefore this map has
tame modulus.\qed
\end{example}

\begin{example}
\label{ex.21}
Let $(W_k)_{k\in\Zb}$ be a given bounded sequence, and consider the map
$z\mapsto X(z):=\left(W_kz_k \right)_{k\in\Zb}$. Its modulus is
$\mod{X}(z):=\left\{|W_k|z_k \right\}_{k\in\Zb}$, which is linear
and bounded as a map from $\Ph_s$ to $\Ph_s$. Therefore, according to
example \ref{linear} it has tame modulus. \qed
\end{example}

\subsection{The theorem}
\label{theorem}

We come now to normal forms.  To define what we mean by normal form we
introduce the complex variables
\begin{eqnarray}
\label{variables}
\xi_{l}:=\frac{1}{\sqrt 2} (p_{l}+\im q_{l})\ ;\ 
\eta_{l}:=\frac{1}{\sqrt 2} (p_{l}-\im q_{l})\quad l\geq 1\ ,
\end{eqnarray}
in which the symplectic form takes the form $\sum_l\im \ d\xi_l\wedge
d\eta_l$. 

\begin{remark}
\label{actio}
In these complex variables the actions are given by
$$
I_j=\xi_j\eta_j\ .
$$\qed
\end{remark}

%

Consider a polynomial function $\Ze$ and write it in the form
\begin{equation}
\label{Z}
\Ze(\xi,\eta)=\sum_{k,l\in\Nn^\Nn}\Ze_{kl}\eta^k\xi^l \ .
\end{equation}

\begin{definition}
\label{norfor}
Fix two positive parameters $\gamma$ and $\alpha$, and a positive
integer $N$. A function $\Ze$ of the form (\ref{Z}) will be said to be in
$(\gamma,\alpha,N)$--normal  form with respect to $\omega$ if $
\Ze_{kl}\not=0 $ implies
\begin{equation}
\label{fornor}
\left|\omega\cdot (k-l)\right|< \frac{\gamma}{N^\alpha} 
\qquad \text{and}\qquad
\sum_{j\geq N+1} k_j+l_j\leq 2\ .
\end{equation}\qed 
\end{definition}

Consider the formal Taylor expansion of $P=P(p,q)$, namely
$$
P=P_3+P_4+...
$$
with $P_j$ homogeneous of degree $j$. We assume

\begin{itemize}
\item[(H)] For any
$s\geq 1$ the vector field $X_P$ is of class $C^{\infty}$ from a
neighbourhood of the origin in $\Ph_s(\R)$ to $\Ph_s(\R)$. Moreover,
for any $j\geq 3$ one has $P_j\in T_M$, namely its Hamiltonian vector
field has tame modulus. \qed
\end{itemize}

Given three numbers 
$R>0$, $r\geq 1$ and $\alpha\in\R$  define the functions
\begin{equation}
\label{nstar}
N_*(r,\alpha,R):=\left[ R^{-(1/2r\alpha)}\right]\ ,\quad s_*(r,\alpha):=2\alpha
r^2 +2\ .
\end{equation}

\begin{theorem}
\label{main}
Let $H$ be the Hamiltonian given by \eqref{hampq}, \eqref{h0pq}, with
$P$ satisfying $(H)$. Fix positive $\gamma$, and $\alpha$.
Then for any $r\geq 1$, and any $s\geq s_*=s_*(r,\alpha)$ there exists
a positive number $R_s$ with the following properties: for any $R<R_s$
there exists an analytic canonical transformation $\Tr_R:B_{s}(R/3)\to
B_{s}(R)$ which puts the Hamiltonian in the form
\begin{equation}
\label{for.nor}
(H_0+P)\circ\Tr_R= \homega+\Ze+\resto
\end{equation}
where $\Ze\in T_M$ is a polynomial of degree at most $r+2$ which is in
$(\gamma,\alpha,N_*)$--normal form with respect to $\omega$, and
$\resto\in C^{\infty}(B_{\R,s}(R/3))$ is small, precisely it fulfills
the estimate
\begin{equation}
\label{stime2}
\sup_{\norma{(p,q)}_s\leq R}\norma{X_{\resto}(p,q)}_s\leq
{C_s} R^{r+\frac32}\ .
\end{equation}
Finally the canonical transformation fulfills the estimate
\begin{equation}
\label{stime}
\sup_{\norma{z}_s\leq R}\norma{z-\Tr_R(z)}_s\leq C_s
R^2\ .
\end{equation}
Exactly the same estimate is fulfilled by the inverse canonical
transformation.  The constant $C_s$ does not depend on $R$.
\end{theorem}

The proof of this abstract theorem is postponed to section \ref{proof}.

Assuming that the frequencies are nonresonant one can easily get
dynamical informations. Precisely, let $r$ be a positive integer, assume

\begin{itemize}
\item[($r$-NR)] There exist $\gamma>0,$ and $\alpha\in\R$ such that
for any $N$ large enough one has
\begin{eqnarray}
\label{nr.1}
\left|\sum_{j\geq 1}\omega_jk_j
\right|\geq\frac{\gamma}{N^\alpha}\ ,
\end{eqnarray}
for any $k\in\Z^\infty$, fulfilling
$0\not=|k|:=\sum_j|k_j|\leq r+2$, $\sum_{j>N}|k_j|\leq2$.\qed
\end{itemize}

\medskip

\begin{remark}\label{rem:NF}
 If the frequency vector satisfies assumption ($r$-NR) then any
polynomial of degree $r+2$ which is in $(\gamma,N,\alpha)$-normal form
with respect to $\omega$ depends only on the actions $\xi_{j}\eta_{j}=
\frac{1}{2}(p_{j}^{2}+q_{j}^{2})$, $j\geq 1$. \qed
\end{remark}

We thus have the following 

\begin{corollary}
\label{diri}
Fix $r$, assume (H,$r$-NR), and consider the system $H_0+P$, then the
normal form $\Ze$ depends on the actions only.
\end{corollary}

In this case one has 

\begin{corollary} 
\label{3.9} 
Fix $r$, assume (H,$r$-NR), and consider the system $H_0+P$.  For any
$s$ large enough there exists $\epsilon_s$ and $c$ such that if
the initial datum belongs to $\Ph_s$ and fulfills
\begin{equation}
\label{d}
\epsilon:=\|z(0)\|_s<\epsilon_s
\end{equation}
one has
\begin{itemize}
\item[(i)]$\displaystyle{\norma{z(t)}_s\leq 2\epsilon}\ ,\quad
\text{for}\quad |t|\leq\frac c{\epsilon^{r+1/2}}$

\item[(ii)] $\displaystyle{
\left|I_j(t)-I_j(0)\right|\leq\frac{\epsilon^3}{j^{2s}}} ,\quad
\text{for}\quad |t|\leq\frac c{\epsilon^{r+1/2}}$

\item[iii)] There exists a torus $\T_0$ with the following properties:
  For any $s_1<s-1/2$ there exists $C_{s_1}$ such that 
\begin{equation}
\label{ds1}
d_{s_1}(z(t),\T_0)\leq C_{s_1} \epsilon^{\frac{r_1}2+1}\ ,\quad
\text{for}\quad |t|\leq\frac 1{\epsilon^{r-r_1+\frac12}}
\end{equation}
where $r_1\leq r$ and $d_{s_1}(.,.)$ is the distance in $\Ph_{s_1}$.
\end{itemize}
\end{corollary}
\noindent {\bf Proof.} Define $R:=8\epsilon$ and use theorem
\ref{main} to construct the normalizing canonical transformation
$z=\Tr_R(z')$.  Denote by $I_j'$ the actions expressed in the
variables $z'$. Define the function $\Nc(z'):=\norma{z'}_s^2\equiv
\sum_j j^{2s}I'_j$. By
(\ref{stime}) one has $\N(z'(0))\leq R^2/ 62$ (provided $R$,
i.e. $\epsilon$ is small enough).  One has
$$
\frac{d\Nc}{dt}(z')=\poisson{\resto}{\Nc}(z')
$$  
and therefore, as far as $\Nc(z'(t))< R^2/9$, one has 
\begin{equation}
\label{norma2}
\left|\frac{d\Nc}{dt}(z')\right|\leq C R^{r+5/2}=C'\epsilon^{r+5/2}\ .
\end{equation}
Denote by $T_f$ the escape time of $z'$ from $B_s(R/3)$.  Remark that
for all times smaller than $T_f$, (\ref{norma2}) holds. So one has
$$
\frac{R^2}{9}=\N(z'(T_f))\leq \Nc(z'(0))+ C' \epsilon^{r+5/2} T_f
$$
which (provided the constants are chosen suitably) shows that $T_f>
C\epsilon^{-(r+1/2)}$. Going back to the original variables one gets
the estimate (i).  To come to the estimate (ii) just remark that  
$$
|I_j(t)-I_j(0)|\leq
 |I_j(t)-I'_j(t)|+|I'_j(t)-I'_j(0)|+|I'_j(0)-I_j(0)| 
$$
and that $j^{2s}I_j$ is a smooth function on $\Ph_s$ and therefore,
using (\ref{stime}) together with (i), it is easy to estimate the
first and the last terms at r.h.s. The middle term is estimated by
computing the time derivative of $j^{2s}I'_j$ with the Hamiltonian and
remarking that its time derivative is of order $\epsilon^{r+5/2}$. 

Denote by $\bar I_j:=I_j(0)$ the initial actions, in the normalized
coordinates. Up to
the considered times 
\begin{equation}
\label{delI}
\left|I_j(t)-\bar I_j\right|\leq\frac{C\epsilon^{2r_1}}{j^{2s}}\ .
\end{equation}
Define the torus
$$
\T_0:=\left\{z\in\Ph_s\ :\ I_j(z)=\bar I_j\ ,j\geq 1 \right\}
$$
One has
\begin{equation}
\label{de2}
d_{s_1}(z(t),\T_0)\leq \left[\sum_{j}j^{2s_1}\left| \sqrt{I_j(t)}-\sqrt{\bar
  I_j} \right|^2 \right]^{1/2}
\end{equation}
Notice that for $a,b\geq 0$ one has,
$$
\left|\sqrt{a}-\sqrt{b}\right|\leq \sqrt{\va{a-b}}\ .
$$ 
Thus, using
(\ref{delI}), one has that
$$
\left[d_{s_1}(z(t),\T_0) \right]^2\leq \sum_{j}\frac{j^{2s}|I_j(t)-\bar
  I_j|}{j^{2(s-s_1)}}\leq \sup{j^{2s}|I_j(t)-\bar
  I_j|}\sum\frac{1}{j^{2(s-s_1)}}
$$
which is convergent provided $s_1<s-1/2$ and gives iii). \qed

\begin{remark}
\label{tau}
It can be shown that, if the vector field of the nonlinearity has tame
modulus when considered as a map from $\Ph_s$ to $\Ph_{s+\tau}$ with
some positive $\tau$ (as in the case of the nonlinear wave equation),
then one can show that both the vector field of $\Ze$ and of $\resto$
are regularizing, in the sense that they map $\Ph_s$ into
$\Ph_{s+\tau}$. Moreover the estimates
(\ref{stime2}), (\ref{stime}), are substituted by     
\begin{eqnarray}
\label{stime3}
\sup_{\norma{(p,q)}_s\leq R}\norma{X_{\resto}(p,q)}_{s+\tau}\leq
{C_s} R^{r+\frac32}\ .
\\
\label{stime4}
\sup_{\norma{z}_s\leq R}\norma{z-\Tr_R(z)}_{s+\tau}\leq C_s
R^2\ ,
\end{eqnarray}
and the estimate (\ref{ds1}) holds with
$s_1<s+(\tau-1)/2$. \qed
\end{remark}

%
%

\section{Applications}
\label{appli}

\subsection{An abstract model of Hamiltonian PDE}
\label{abstract model}

In this section we present a general class of Hamiltonian PDEs to which
 theorem \ref{main} applies. We focus only on the one dimensional
 case but the discussion of the tame modulus property can be easily
 generalized to higher dimension.  

Denote by $\T$ the torus, $\T := \R / 2\pi \Z$ and consider the space
$L^2(\T)\times L^2(\T)$ endowed by the symplectic form
$$ 
\Omega \left( (p_{1},q_{1}),(p_{2},q_{2}) \right) := \langle J (
p_{1} , q_{1} ) ; ( p_{2} , q_{2} ) \rangle_{L^{2}\times L^{2}}
$$
where
$$
J\left( p ,q  \right) =
\left( -q , p  \right)
$$
With this symplectic structure the Hamilton equations associated to 
a Hamiltonian function,  $H:L^2\times L^2\supset D(H)\to \R$,
 read
$$
\left\{ \begin{array}{c} \dot p = -\nabla_{q }H \\ \dot q = \nabla_{p}H
\end{array} \right.\ 
$$
with $\nabla _q$ and $\nabla _p$ denoting the $L^2$ gradient with
respect to the $q$ and the $p$ variables respectively.

Let $A$ be a self-adjoint operator on $L^2 (\T)$ with pure point
spectrum $(\omega_j)_{j\in\Z}$. Denote by $\varphi_j$, $j\in\Z$, the
associated eigenfunctions, i.e.
$$
A\varphi_j=\omega_j\varphi_j\ .
$$
The sequence $(\varphi_j)_{j\in \Z}$ defines a Hilbert basis of
$L^2(\T)$. 

We use this operator to define the quadratic part of the
Hamiltonian. Precisely we put 
\begin{eqnarray}
\label{H0conc}
H_0&:=&\frac{1}{2} \left(\left\langle Ap,p \right\rangle_{L^2} +\left\langle
Aq,q \right\rangle_{L^2} \right)
\\
&=&\sum_{j\in\Z}\omega_j\frac{p_j^2+q_j^2}{2}
\end{eqnarray}
where $q_j$ is the component of $q$ on $\varphi_j$ and similarly for
$p_j$. Remark that here the indexes run in $\Z$, so that the set
$\Zb$ has to be substituted by the disjoint union of two copies of
$\Z$.

Concerning the normal modes $\varphi_j$ of the quadratic part, we  assume 
they are well localized with respect to the exponentials:
Consider the Fourier expansion of $\varphi_j$,
$$
\varphi_{j}(x) = \sum_{k\in \Z} \varphi_{j}^k e^{\im kx}\ ,
$$ 
we assume
\begin{itemize}
\item[(S1)] For any $n>0$ there exists a constant $C_n$ such that for
all $j\in \Z$ and $k\in \Z$
\begin{equation}
\label{exploc}
\Va{\varphi_{j}^k} \leq \max_{\pm}\frac{C_n}{(1+|k\pm j|)^n} \ .
\end{equation}\qed
\end{itemize}

\begin{example}
\label{ex.3}
If $A=-\partial_{xx}$, then $\varphi_{j}(x)=\sin jx$ for $j>0$ and
$\varphi_j(x)=\cos(-jx)$ for $j\leq 0$, is well localized with respect
to the exponentials.\qed
\end{example}

\begin{example}
\label{ex.5}
 Let $A=-\partial_{xx}+V$, where $V$ is a $C^{\infty}$, $2\pi$
 periodic potential.  Then $\varphi_{j}(x)$ are the eigenfunctions of
 a Sturm Liouville operator. By the theory of Sturm Liouville
 operators (cf \cite{Mar86,PT}) it is well localized with respect to
 the exponentials (cf \cite{CW92,KaMi01}). Moreover, if $V$ is even
 then one can order the eigenfunctions in such a way that $\varphi_j$
 is odd for strictly positive $j$ and even for negative $j$.\qed
\end{example}

\begin{remark}
\label{h^s}
Under assumption (S1) the space $H^s(\T)$ coincides with the space of
the functions $q=\sum_{j}q_j\varphi_j$ with $(q_j)\in\ell^{2}_s$.\qed
\end{remark}

On the symplectic space $\B_{s} :=H^{s}(\T) \times H^{s}(\T)$ consider
the Hamiltonian system, $H= H_0+P$ with $H_0$ defined by
(\ref{H0conc}) and $P$ a $C^{\infty}$ function which has a zero of
order at least three at the origin.

   While it is quite hard to verify the tame modulus property
when the basis $\varphi_j$ is general, it turns out that it
is quite easy to verify it using the basis of the complex
exponentials. So it is useful to reduce the general case to that of
the Fourier basis. Let $\Phi$ be the isomorphism between
 $\Ph_s$ and  $\B_s$ given by
\begin{equation}
\label{isom}
\Ph_s\ni(p_k,q_k)\mapsto  \Phi(p_k,q_k):= (\sum_kp_ke^{\im k x},
\sum_kq_ke^{\im k x})
\end{equation}
Then any polynomial vector field on $\B_s$ induces a polynomial vector field on
$\Ph_s$.

\begin{definition}
\label{expbas}
A polynomial vector field $X:\B_s\to\B_s$ will be said to have {\it
tame modulus with respect to the exponentials} if   the polynomial 
vector field $\Phi^{-1} X \Phi$ has tame modulus.\qed
\end{definition}

\begin{example}
\label{concreto}
By examples \ref{ex.2}, \ref{ex.21} and the result of lemma
\ref{compo}  ensuring that the composition of maps having tame modulus
has tame modulus, one has that given a $C^\infty$ periodic function
$g$, the Hamiltonian functions
\begin{eqnarray}
\label{ese}
H_1(p,q)=\int_{\T} g(x)p(x)^{n_1}q(x)^{n_2} dx\ ,
\\
H_2(p,q)=\int_{\T\times\T}
p(x)^{n_1}q(x)^{n_2}g(x-y)p(y)^{n_3}q(y)^{n_4}dxdy    
\end{eqnarray}
have a vector field with tame modulus with respect to the
exponentials. \qed
\end{example}

The main result we will use to verify the tame modulus property is 

\begin{theorem}
\label{prop:tame}
Let $X:\B_s\to\B_s$ be a polynomial vector field having tame modulus
with respect to the exponentials; assume (S1), then $X$ has tame
modulus.
\end{theorem}
The 
proof is detailed in appendix \ref{ApT}.

\begin{example}
\label{concretobis}
  By example \ref{concreto} given a $C^\infty$ periodic function $g$
  and a $C^\infty$ function $f$ form  $\R^2$ or
  $\R^{4}$ into $\R$, the Hamiltonian functions
\begin{eqnarray}
\label{esebis}
P_1(p,q)=\int_{\T} g(x)f(p(x),q(x)) dx\ ,\quad
\\
P_2(p,q)=\int_{\T\times\T} g(x-y)f(p(x),p(y),q(x),q(y))dxdy   
\end{eqnarray}
satisfy hypothesis $(H)$.\qed
\end{example}

\begin{remark}
\label{cou}
All the above theory extends in a simple way to the case where the
space $H^s(\T)$ is substituted by $H^s(\T)\times H^s(\T)...\times
H^s(\T)$, or by $H^{s}(\T^d)$, cases needed to deal with equations in
higher space dimensions and systems of coupled partial differential
equations.
\end{remark}

To deal with Dirichlet boundary conditions, we will consider the space

\begin{equation}
\label{Hs}
\Hc_s:=\text{Span}\left((\varphi_j)_{j\geq 1} \right)\ .
\end{equation}
 
\begin{example}
\label{ex.4}
Consider again examples \ref{ex.3}, and \ref{ex.5} with $V$ an even
function. In these cases the function space $\mathcal H_{s}$ is the
space of the functions that extend to $H^s$ skew--symmetric periodic
functions of period $2\pi$.  Equivalently $\mathcal H_s$ is the
space of the functions $q\in H^s([0,\pi])$ fulfilling the
compatibility conditions
\begin{equation}
\label{comp}
q^{(2j)}(0)=q^{(2j)}(\pi)=0\ ,\quad 0\leq j\leq \frac{s-1}{2}\ , 
\end{equation}
i.e. a generalized Dirichlet condition. Actually, recall that for $V$
even, the Dirichlet eigenvalues of $A=-\partial_{xx}+V$ are periodic
eigenvalues.  Thus, due to our choice of the labeling, the sequence of
eigenvalues $(\omega_{j})_{j\geq 1}$ corresponds to the Dirichlet
spectrum of $A$ and $(\varphi_{j})_{j\geq 1}$ are the corresponding
Dirichlet eigenfunctions.\qed
\end{example}

Concerning the tame modulus property, one has in this case

\begin{corollary}
\label{diri2}
Assume that the subspace $\Hc_s\times\Hc_s\subset \B_s$
is mapped smoothly into itself by the nonlinearity $X_P$, then 
  if $H_0+P$ fulfills assumption (H) in $\B_s$ then  the
  restriction of $H_0+P$ to  $\Hc_s\times\Hc_s$ fulfills assumption
  (H). 
\end{corollary}

\subsection{Nonlinear wave equation}
\label{NLW}

As a first concrete application 
we consider a nonlinear wave equation
\begin{eqnarray}
\label{nlw}
u_{tt}-u_{xx}+V(x)u=g(x,u)\ ,\quad x\in\T \ , \ t\in \R \ ,
\end{eqnarray}
where $V$ is a $C^\infty$, $2\pi$ periodic potential and $g\in
C^{\infty}(\T\times\U)$, $\U$ being a neighbourhood of the origin in
$\R$.

Define the operator $A:=(-\partial_{xx}+V)^{1/2}$, and introduce the
variables $(p,q)$ by
$$
q:=A^{1/2}u\ ,\quad p:= A^{-1/2}u_t\ ,
$$
then the Hamiltonian takes the form $H_0+P$, with $H_0$ of the form
(\ref{H0conc}) and $P$ given by
\begin{equation}
\label{PNLW}
P(q)=\int_{\T}G(x,A^{-1/2}q)dx\sim \sum_{j\geq
  3}\int_{\T}G_j(x)(A^{-1/2}q)^j dx
\end{equation}
where $G(x,u)\sim \sum_{j\geq 3}G_j(x)u^j$ is such that
$\partial_{u}G=-g$ and $\sim$ denotes the fact that the r.h.s. is the
asymptotic expansion of the l.h.s.

The frequencies are the square roots of the eigenvalues of the
Sturm--Liouville operator
\begin{equation}
\label{ST}
-\partial_{xx}+V
\end{equation}
and the normal modes $\varphi_j$ are again the
eigenfunctions of (\ref{ST}). In particular, due to example \ref{ex.5}
they fulfill (S1). 

\begin{proposition}
\label{TNLW}
The nonlinearity $P$ has tame modulus.
\end{proposition} 
\proof Denote by $P_j$ the $j$-th Taylor coefficient of $P$, then
$$
X_{P_j}=\left(-A^{-1/2}G_j(x)(A^{-1/2}q )^j,0\right)
$$ 
is the composition of three maps, namely 
$$
q\mathop{\mapsto}^1A^{-1/2}q\mathop{\mapsto}^2 G_j(x)(A^{-1/2}q )^j
\mathop{\mapsto}^3 - A^{-1/2}G_j(x)(A^{-1/2}q )^j
$$
The first and the third ones are smoothing linear maps, which therefore
have tame modulus, and the second one has tame modulus with respect
to the exponentials. By lemma \ref{compo} the thesis follows.\qed 

Thus the system can be put in ($\gamma,\alpha,N_*$)--normal form. To
deduce dynamical informations we need to know something on the
frequencies.

\subsubsection{Dirichlet boundary conditions}
\label{dirichlet}

First remark that if both $V$ and $G(x,u)$ are even, then the
eigenfunctions and the eigenvalues can be ordered according to example
\ref{ex.4}, and moreover the space $ \Hc_s\times\Hc_s $ is invariant
under $X_{P}$, so that assumption (H) holds also for the system with
Dirichlet boundary conditions.  The nonresonance condition ($r$-NR) is
satisfied for almost all the potentials in the following sense: Write
$V=V_0+m$, with $V_0$ having zero average. Let $\lambda_j$ be the
sequence of the eigenvalues of $-\partial_{xx}+V_0$, then  the
frequencies are 
\begin{equation}
\label{frewave}
\omega_j:=\sqrt{\lambda_j+m}
\end{equation}
Let $m_0:=\min_j\lambda_j$, then the following theorem holds

\begin{theorem}
\label{nres} 
Consider the sequence $\{\omega_j\}_{j>0}$ given by \eqref{frewave},
for any $\Delta>m_0$ 
there exists a set $\I\subset (m_0,\Delta)$ of measure $\Delta-m_0$ such
that, if $m\in \I$ then for any $r\geq 1$  assumption
($r$-NR) holds.  
\end{theorem}
Theorem \ref{nres} was proved in ref. \cite{Bam03}; in section
\ref{B} we will reproduce the main steps of the proof.

So, in the case of Dirichlet boundary, conditions it is immediate to
conclude that for $m$ in the set $\I$ corollary \ref{3.9}
applies to the equation \eqref{nlw}. Moreover it is easy to verify
that we are in the situation of remark \ref{tau} with $\tau=1$ and
therefore  (\ref{ds1}) holds for $s_{1}<s$.

\begin{remark}
\label{nlwbam}
Such a result was already obtained in \cite{Bam03}.
\end{remark}

\subsubsection{Periodic boundary conditions}
\label{periodic}

In
the case of periodic boundary conditions the frequencies are again of
the form \eqref{frewave} with $\lambda_j$ being the periodic eigenvalues
of the operator $-\partial_{xx}+V_0$. We label them in such a way
that, for  $j>0$, $\lambda_j$ are the Dirichlet eigenvalues and,
for  $j\leq 0$, $\lambda_j$ are the Neumann eigenvalues. 

The situation is more complicated than in the Dirichlet case since
asymptotically $\omega_{j} \sim \omega_{-j}$ and we cannot hope
condition ($r$-NR) to be satisfied. Actually, for analytical $V$ one
has
$$
\va{\omega_{j}- \omega_{-j}}\leq C e^{-\sigma \vaj}
$$ and thus $\va{\omega_{j} - \omega_{-j}}$ cannot be bounded from
below by $1/N^{\alpha}$ as soon as $\vaj \geq C \ln N$. The
forthcoming theorem essentially states that for typical small $V$ this
is the only case where condition ($r$-NR) is not satisfied.

Consider a potential $V$ of the form 
\begin{equation}
\label{V1}
V(x)=m+\sum_{k\geq 1}v_k \cos kx
\end{equation}
we will use the values $(v_k)_{k\geq 1}$ and the value of the mass $m$ as random
variables. More precisely, having fixed a positive $\Delta$
and a positive $\sigma$, for any $R>0$ we consider the probability
space
\begin{equation}
\label{ps1}
\V_R:=\left\{(m,(v_k)_{k\geq 1})\ :\ m':=\Delta^{-1}m \in[0,1]\ ,\
v'_k:=R^{-1}e^{\sigma k}v_{k}\in[-\frac{1}{2},\frac{1}{2}]  \right\}
\end{equation}
endowed by the product probability measure on $(m',v'_{k})$. We will identify
$V$ with the coefficients $(m,v_k)$. 

\begin{theorem}
\label{teodiofper}
For any positive $r$ there exist a positive $R$ and a set $\Scr\subset \V_{R}$
such that
\begin{itemize}
\item[i)] for any $V\in\Scr$ there exists a positive $\gamma$,  a
positive $\alpha$ and a  positive $b$ such that for any $N\geq 1$
\begin{eqnarray}
\label{nr.2}
|\sum_{j\in\Z}\omega_j
k_j|\geq\frac{\gamma}{N^\alpha}\ , 
\end{eqnarray}
for any $k\in\Z^{\Z}$, fulfilling $0\not=|k|\leq r+2$, $\sum_{|j|>
  N} |k_j|\leq 2$
except  if \\
$$
(k_{j} =0 \text{ for  } \vaj \leq b \ln N)\  
\text{and } \ (k_{j}+k_{-j}=0\  \text{ for }\vaj>b \ln N)\ .
$$
\item[ii)] $|\V_{R}-\Scr|=0$, where $|.|$ denotes the measure of its
  argument. 
\end{itemize}
\end{theorem}
The proof is postponed to section \ref{periodic wave}.

As the assumption ($r$-NR) is no more satisfied, corollary \ref{diri} does 
not apply. However one has
\begin{lemma}
\label{per.nls} Assume $V\in \Scr$.
Let $\xi^k\eta^l$ be a monomial in $(\gamma,\alpha,N)$--normal form
for the system with periodic boundary conditions, then one has 
\begin{eqnarray}
\label{per.3}
\poisson{\xi_j\eta_j}{\xi^k\eta^l}=0\ ,\quad \text{for } \vaj \leq 
b\ln N 
\\
\label{per.4}
\poisson{\xi_j\eta_j+\xi_{-j}\eta_{-j} }{\xi^k\eta^l}=0\ ,\quad
\text{for } \vaj > 
b\ln N
\end{eqnarray}
\end{lemma}

\noindent
{\it Proof.} Denote $J:=b\ln N$ and
  $\gamma_j:=\omega_j-\omega_{-j}$. Assume that $b$ is so large that
  $|\gamma_j|<\gamma/2N^\alpha$ $\forall j>J$.  Let $\xi^k\eta^l$ be
  in $(\gamma,\alpha,N)$--normal form; denote $K:=(k-l)\in
  \Z^{\Z}$. By definition of normal form one has
$$ 
\frac{\gamma}{N^\alpha}>\left|\sum_{j\in\Z}\omega_jK_j
\right|= \left| \sum_{\vaj\leq J}\omega_jK_j
+\sum_{j>J}\omega_j(K_j+K_{-j}) -\sum_{j>J}\gamma_jK_j\right|
$$
from which 
$$
\left| \sum_{\vaj\leq J}\omega_jK_j
+\sum_{j>J}\omega_j(K_j+K_{-j})\right|<\frac{\gamma}{2N^\alpha}
$$ 
then, by theorem \ref{teodiofper} one has $K_j=0$ for $\vaj\leq J$
and $K_j+K_{-j}=0$ for $j>J$. As a consequence the normal form
commutes with $I_j$ for all $\vaj\leq J$. Write 
$$
\xi^k\eta^l=\prod_{j\in\Z}\xi_j^{k_j}\eta_j^{l_j}=
\left(\prod_{\vaj\leq J}\xi_j^{k_j}\eta_j^{l_j}\right)
\prod_{j>J}\xi_j^{k_j}\eta_j^{l_j} 
\xi_{-j}^{k_{-j}}\eta_{-j}^{l_{-j}}  
$$
 Compute now   
\begin{eqnarray*}
&&\poisson{\xi_j\eta_j+\xi_{-j}\eta_{-j} }{\xi^k\eta^l}
\\
&=& \left(\prod_{\vaj\leq J}\xi_j^{k_j}\eta_j^{l_j}\right)
\left(\prod_{n>J,n\not=j}\xi_n^{k_n}\eta_n^{l_n}
\xi_{-n}^{k_{-n}}\eta_{-n}^{l_{-n}} \right)
\\&&
\poisson{\xi_j\eta_j+\xi_{-j}\eta_{-j}}{  \xi_j^{k_j}\eta_j^{l_j}
  \xi_{-j}^{k_{-j}}\eta_{-j}^{l_{-j}} } 
\\
&=&\left(\prod_{\vaj\leq J}\xi_j^{k_j}\eta_j^{l_j}\right)
\left(\prod_{n>J,n\not=j}\xi_n^{k_n}\eta_n^{l_n}
\xi_{-n}^{k_{-n}}\eta_{-n}^{l_{-n}} \right)
\\
&&
(-{\rm i})\left[(l_j-k_j)+(l_{-j}-k_{-j})\right] {  \xi_j^{k_j}\eta_j^{l_j}
  \xi_{-j}^{k_{-j}}\eta_{-j}^{l_{-j}} }
\\
&=&-{\rm i} (K_j+K_{-j}) \xi^k\eta^l =0\ .
\end{eqnarray*}
\qed

This allows to get dynamical consequences. Define
$J_j:=I_j+I_{-j}$ then one has

\begin{theorem}
\label{thm.nlwper}
Consider the wave equation \eqref{nlw} with periodic boundary
conditions, fix $r$, assume $V\in \Scr$. For any $s$ large enough,
there exists $\varepsilon_{s} >0$ and $C_{s}>0$ such that if the
initial datum $(u_0,\dot u_0)$ belongs to $H^s(\T)\times H^{s-1}(\T) $
and fulfills $\varepsilon := \norma{u_0}_{s}+\norma{\dot u_0}_{s-1} <
\varepsilon_{s}$ then
$$
\norma{u(t)}_{s}+\norma{\dot u(t)}_{s-1} \leq 2 \varepsilon
\text{ for all } \va{t}\leq C_{s}\varepsilon^{-r}\ .
$$
Further there exists $C'_{s}$ such that for all
$\va{t}\leq C_{s}\varepsilon^{-r}$ one has
\begin{eqnarray*}
&\va{I_{j}(t) -I_{j}(0)} \leq \displaystyle{\frac{1}{|j|^{2s}}}
\varepsilon^{3}\quad \text{ for } \vaj \leq -C'_{s}\ln \varepsilon \\
&\va{J_{j}(t) -J_{j}(0)} \leq \displaystyle{\frac{1}{|j|^{2s}}}
\varepsilon^{3} \quad \text{ for } j > -C'_{s}\ln \varepsilon \ .
\end{eqnarray*}
\end{theorem}
Roughly speaking, the last property means that energy transfers are
allowed only between modes of index $j$ and $-j$ with $j$ large.

\subsection{NLS in 1 dimension}
\label{s.NLS}

We will consider here only the case of Dirichlet boundary
conditions. However, to fit the scheme of section \ref{abstract model}
we start
with the periodic system. 
 
Consider the nonlinear Schr\"odinger equation 
\begin{eqnarray}
\label{NLS}
-\im \dot \psi=-\psi_{xx}+V\psi+\frac{\partial
  g(x,\psi,\psi^*)}{\partial \psi^*}\ ,\quad x\in \T,\quad t\in \R
\end{eqnarray}
where $V$ is a $C^\infty$, $2\pi$ periodic potential. We assume that
$g(x,z_{1},z_{2})$ is $C^\infty(\T\times\U)$, $\U$ being a
neighbourhood of the origin in $\C\times \C$. We also assume that $g$
has a zero of order three at $(z_{1},z_{2})=(0,0)$ and that
$g(x,z,z^{*})\in \R$.

The Hamiltonian function of the system is 
\begin{equation}
\label{NLSH}
H=\int_{-\pi}^{\pi}\frac{1}{2}\left(|\psi_x|^2+V|\psi|^2
\right) 
+g(x,\psi(x),\psi^*(x))dx
\end{equation}

Define $p$ and $q$ as the real and imaginary parts of $\psi$,
namely write $ \psi=p+\im q $. Then the operator $A$ is the
Sturm--Liouville operator $-\partial_{xx}+V$ with periodic boundary
conditions, the frequencies $\omega_j$  are 
the corresponding
eigenvalues and the normal modes $\varphi_j$  are 
the corresponding eigenfunctions. 

Then property (S1) is a consequence of Sturm Liouville theory (see
example \ref{ex.5}) and it is easy to verify that property $(H)$ holds
for the periodic system. To deal with Dirichlet boundary conditions we
have to ensure the invariance of the space 
$\Hc_{s}$ under the vector field of the equation (cf. corollary 
\ref{diri2}). To this end we 
 assume
\begin{equation}
\label{sym}
V(x)=V(-x)\ ,\quad g(-x,-z,-z^{*}) = g(x,z, z^{*})\ .
\end{equation}
Then  (H) holds true  in the Dirichlet context and
thus theorem \ref{main} applies and the Hamiltonian \ref{NLSH} can be put in
$(\gamma,\alpha,N)$- normal form in $\Hc_{s}\times \Hc_{s}$.
We are going to prove that for
typical small $V$ such a normal form is integrable.

 Fix $\sigma>0$ and, for any positive $R$ define the space of the
 potentials, by
\begin{equation}
\label{Vnls}
\V_R:=\left\{V(x)= \sum_{k\geq1} v_k\cos kx\mid 
v'_k:=R^{-1}e^{\sigma k}
\in\left[-\frac{1}{2},\frac{1}{2} \right]
 \mbox{ for }k\geq 1 \right\}
\end{equation}
that we endow with the product probability measure. We remark that any
potential in $\V_R$ has size of order $R$, is analytic and has zero
average. We also point out that the choice of zero average was done
for simplicity since the average does not affect the resonance
relations among the frequencies.

\begin{theorem}
\label{NLSdir}
For any $r$ there exists a positive $R$ and a set $\Sc\subset \V_R$
such that property ($r$-NR) holds for any potential $V\in\Sc$ and
$\left|\V_R-\Sc \right|=0$.
\end{theorem}

The proof of this theorem is postponed to section \ref{B}.

Thus corollary \ref{3.9} holds and every small amplitude solution
remains small for long times and approximatively lies on a finite
dimensional torus.

\begin{remark}
\label{hartree}
We remark that the theory applies also to Hartree type equations of
the form
\begin{equation}
\label{har1}
-\im \psi_t=-\psi_{xx}+V\psi+(W*|\psi|^2)\psi\ .
\end{equation}
\end{remark}

\subsection{Coupled NLS in 1 dimension}
\label{coupledNLS}

As an example of a system of coupled partial differential equations we
consider a pair of NLS equations. 
From the mathematical point of view the interest of this example is
that, since it does not have a positive
definite energy, nothing is a priori known about global
existence of its solutions in any phase space. So, consider the
Hamiltonian
\begin{equation}
\label{cou.1}
H=\int_{\T}\left|\psi_x\right|^2-\left|\phi_x\right|^2+V_1
\left|\psi\right|^2 -V_2
\left|\phi\right|^2 -g(x,\psi,\psi^*,\phi,\phi^*)
\end{equation}
The corresponding equations of motion have the form 
\begin{eqnarray}
\label{cou.2}
-{\rm i}\dot \psi&=& -\psi_{xx}+V_1\psi-\partial_{\psi^*}g
\\
\label{cou.3}
{\rm i}\dot \phi&=& -\phi_{xx}+V_2\phi+\partial_{\phi^*}g
\end{eqnarray}
Assume as in the previous sections that the potentials and the
function $g$ are of class $C^\infty$, then condition (H) holds for the
system with periodic boundary conditions. Assuming also that the
potentials $V_1,$ $ V_2$ and $g$ are even in each of the variables,
one has that the space of odd functions is invariant and therefore the
system with Dirichlet boundary conditions fulfills also condition (H).

We concentrate now on the case of Dirichlet boundary conditions.

The frequencies are given by
$$
\omega_{j}:=\lambda_j^1\ ,\quad \omega_{-j}:=-\lambda_j^2\ ,\quad j\geq 1
$$
where $\lambda_j^1$, and $\lambda_j^2$, are the Dirichlet eigenvalues of
$-\partial_{xx}+V_1$ and $-\partial_{xx}+V_2$ respectively.

Assume that the
potentials vary in the probability space obtained by taking two copies
of the space (\ref{Vnls}). As in the case of periodic NLW (cf. 
subsection \ref{periodic}), $\omega_{j} \sim \omega_{-j}$ and we cannot hope
condition ($r$-NR) to be satisfied. Actually one has
$\va{\omega_{j}-\omega_{-j}}\leq C/\vaj^2$ and, by adapting the proof of theorem
\ref{NLSdir} in the spirit of theorem \ref{teodiofper}, one gets (the 
proof is postponed to subsection \ref{couple}) 

\begin{theorem}
\label{cou.9}
For any positive $r$ there exist a positive $R$ and a set
$\Sc_{\psi,\phi}\subset \V_{R}\times\V_R$ 
such that
\begin{itemize}
\item[i)] for any $(V_1,V_2)\in\Sc_{\psi,\phi}$ there exists a
positive $\gamma$, a positive $\alpha$ and a positive $C$ such that
for any $N\geq 1$
\begin{eqnarray}
\label{nr.21}
|\sum_{j\in\Zb}\omega_j
k_j|\geq\frac{\gamma}{N^\alpha}\ , 
\end{eqnarray}
for any $k\in\Z^{\Z}$, fulfilling $0\not=|k|\leq r+2$, $\sum_{|j|>
  N} |k_j|\leq 2$
except  if \\
$$
(k_{j} =0 \text{ for  } \vaj \leq C  N^{\sqrt{2\alpha}})\  
\text{and } \ (k_{j}+k_{-j}=0\  \text{ for }\vaj>C N^{\sqrt{2\alpha}})\ .
$$
\item[ii)] $|\V_{R}\times\V_R-\Sc_{\psi,\phi}|=0$.
\end{itemize}
\end{theorem}
For the proof see section \ref{couple}.

In particular one deduces long time existence of solutions and long
time stability of the zero equilibrium point:
\begin{theorem}
\label{coul} 
Consider the system (\ref{cou.2},\ref{cou.3}) and fix $r\geq 1$. Provided
$V_1\times V_2\in \Sc_{\psi,{\phi}} $, for any $s$ large enough there
exists a positive $\epsilon_s$ such that if the initial datum
$(\psi_0,\phi_0)\in H^s(\T)\times H^s(\T)$ fulfills
$$ 
\epsilon:=\norma{\psi_0}_{H^s}+\norma{\phi_0}_{H^s}\leq \epsilon_s
$$
then the corresponding solution exists until the time $\epsilon^{-r}$
and fulfills
$$
\norma{\psi(t)}_{H^s}+\norma{\phi(t)}_{H^s}\leq 2\epsilon
$$
\end{theorem}

\subsection{NLS in arbitrary dimension}
\label{dnls}

Consider the following non linear
Schr\"odinger equation in dimension $d\geq 1$
\begin{equation}
\label{NLS.d}
    -\im\psi _{t} = -\Delta \psi + V\ast \psi +\frac{\partial g(x,\psi ,
    \psi ^{*} )}{\partial \psi ^{*}}\ , \quad x\in [-\pi, \pi]^d, t\in
    \R
\end{equation}
with periodic boundary conditions. 

As in the previous sections we assume that
$g(x,z_{1},z_{2})$ is $C^\infty(\T^d\times\U)$, $\U$ being a
neighbourhood of the origin in $\C\times \C$. We also assume that $g$
has a zero of order three at $(z_{1},z_{2})=(0,0)$ and that
$g(x,z,z^{*})\in \R$.

Fix $m>d/2$ and $R>0$, then the potential $V$ is
chosen in the space $\V$ given by

\begin{equation}\label{pot}
\V =\{ V(x)=\sum_{k\in \Z^d}v_k
e^{\im k\cdot x}
\mid v'_{k}:={v_k}{(1+\va{k})^m}/R \in [-1/2,1/2]
\mbox{ for any }k\in \Z^d \}
\end{equation}
that we endow with the product probability measure. In contrast with
the previous cases, here $R$ is arbitrary (it does not need to be
small). 

In this section we denote $H^s\equiv H^s(\Td ; \C)$ the
Sobolev space of order $s$ on the $d$-dimensional torus $\Td$,
$\Ns{\cdot}$ the usual norm on $H^s$. 
Notice also that in this section all the indexes run in $\Z^d$.

The NLS equation \eqref{NLS.d} is Hamiltonian with Hamiltonian
function given by
$$
H(\psi ,\psi^* ) = \int_{\Td} \left(\va{\nabla \psi }^{2} +(V\ast \psi
)\psi^*  + 
g(x,\psi ,\psi^* ) \right) dx.
$$
It is convenient to introduce directly the variables $\xi,\eta$ by
$$\psi (x)=\left( \frac{1}{2\pi}\right)^{d/2} \sum_{k\in \Z^d} \xi_{k} 
e^{\im k\cdot x}\ ,\quad
 \psi^* (x)=\left( \frac{1}{2\pi}\right)^{d/2} \sum_{k\in \Z^d}
\eta_{k} 
e^{-\im k\cdot x},$$ 
so
the Hamiltonian reads
$$
H(\xi,\eta)=  \sum_{k\in \Z^d} \omega_{k} \xi_{k} \eta_{k} \ +\ 
\int_{\Td} g(x,\psi , \psi^* )  
$$   
where the linear frequencies are given by
$\omega_{k}= \va{k}^{2} + v (k)$.

It is immediate to realize that the nonlinearity has tame modulus so
that theorem \ref{main} applies with an adapted definition of
$(\gamma,\alpha,N)$--normal form. 

Remark that, if $\val =\vaj \to \infty$ then $\omega_{j}-\omega_{l}
\to 0$ as $|l|\to\infty$. Thus property ($r$-NR) is violated. The
following theorem ensures this is the only case where it happens.
\begin{theorem} 
\label{res.d}
There exists a set $\Sc \subset \V$ of measure 1 such that, for any 
$V \in \Sc$ the following property holds.  For any positive $r$ there
exist positive constants $\gamma,\
\alpha,$ such that for any $N\geq 1$
\begin{eqnarray}
\label{nr.22}
|\sum_{j\in\Z^d}\omega_j
k_j|\geq\frac{\gamma}{N^\alpha}\ , 
\end{eqnarray}
for any $k\in\Z^{\Z^d}$, fulfilling $0\not=|k|\leq r+2$, $\sum_{|j|>
  N} |k_j|\leq 2$
except  if \\
$$ (k_{j} =0 \text{ for } \vaj \leq N^{\sqrt{\alpha/m}})\ \text{and }
\ (\sum_{|j|=K} k_{j}=0\ \text{ for all }K>N^{\sqrt{\alpha/m}})\ .
$$
\end{theorem}
For the proof see section \ref{B}.

As a consequence theorem \ref{main} does not allow to eliminate
monomial of the form $I_{k_1}\ldots I_{k_r}\xi_j\eta_{l} $ with large
$|j|$ and $|l|$.  Nevertheless, defining for $M>N^{\sqrt{\alpha/m}}$
$J_{M}=\sum_{\va{k}^{2}=M}I_{k}$, we have

\begin{theorem}
\label{thm.nlsdper}
Consider the equation \eqref{NLS.d}, and assume $V\in
\Sc$. Fix $r\geq 1$, then for any $s$ large enough, there
exist $\varepsilon_{s} >0$ and $C_{s}>0$ such that if the initial
datum $\psi(\cdot,0)$ belongs to $H^s(\T^d)$ and fulfills $\varepsilon
:= \norma{\psi(\cdot, 0)}_{s} < \varepsilon_{s}$ then
$$
\norma{\psi(\cdot,t)}_{s}\leq 2 \varepsilon
\text{ for all } \va{t}\leq C_{s}\varepsilon^{-r}\ .
$$
Furthermore there exists an integer $N\sim \epsilon 
^{-\frac{1}{2r\alpha}}$ (where $\alpha$  is defined in theorem 
\ref{res.d})
such that 
\begin{eqnarray*}
&\va{I_{j}(t) -I_{j}(0)} \leq \displaystyle{\frac{C_s}{|j|^{2s}}}
\varepsilon^{3}\quad \text{ for } \vaj \leq N^{\sqrt{\alpha/m}} 
\\
&\va{J_M(t)-J_{M}(0)} \leq \displaystyle{\frac{C_s}{M^{2s}}}
\varepsilon^{3}\quad \text{ for } M >N^{\sqrt{\alpha/m}} \ .
\end{eqnarray*}
\end{theorem}
Roughly speaking, the last property means that energy transfers are
allowed only among modes having indexes with equal large modulus.

If the nonlinearity does not depend on $x$ something more can be
concluded. To come to this point consider the following

\begin{definition}
\label{moment}
Given a monomial
$\xi_{j_1}....\xi_{j_{r_1}}\eta_{l_1}...\eta_{l_{r_2}}$, its momentum
is defined by $j_1+...+j_{r_1}-(l_1+...+l_{r_2})$.\qed
\end{definition}
It is easy to see that if the function $g$ does not depend on $x$ then
the nonlinearity contains only monomials with zero momentum. Moreover
this property is conserved by our iteration procedure defined in 
section \ref{iterative}. Therefore the
normal form has zero momentum and the following corollary holds:

\begin{corollary}
\label{zero}
If the function $g$ does not depend on $x$ then the normal form $\Ze$ of the
system depends on the actions only. 
\end{corollary}
\proof By theorem \ref{res.d} the only non integrable term that the
normal form may contain are of the form 
$$
I_{k_1}...I_{k_l}\xi_j\eta_i
$$ 
with $|i|=|j|$, but the momentum $j-i$ of such a term must vanish and
therefore one must have $i=j$.\qed

Thus the following theorem holds

\begin{theorem} 
\label{thm.nlsd}
    Consider the $d$-dimensional NLS equation \eqref{NLS.d} with
periodic boundary conditions and $g$ independent of $x$.  Assume $V\in
\Sc$.  Fix $r$, then, for $s$ large enough, there exist
$\varepsilon_{s}>0$ and $c_{s}>0$ such that the following properties
hold :

If $\psi(t)$ is the solution of the Cauchy problem \eqref{NLS.d} with 
initial datum $\psi _{0}\in H^s$ satisfying $\varepsilon
:=\Ns{\psi _{0}}\leq
\varepsilon_{s}$ then for all
$$\va{t} \leq 
\frac{c_{s}}{\varepsilon^{r}}$$
the solution $\psi$ satisfies 
$$
\Ns{\psi (t)}\leq 2 \varepsilon \ , \quad \va{I_{k}(t)-I_{k}(0)}\leq 
\frac{1}{\va{k}^{2s}}\varepsilon^{3}\ .
$$
\end{theorem}

One also has that corollary \ref{3.9} applies and therefore any
initial datum which is smooth and small enough give rise to a solution
which is $\epsilon^{r_1}$ close to a torus up to times
$\epsilon^{-r_2}$. 

The results of this section were
announced in \cite{BG}.

\section{Proof of the Normal Form Theorem}
\label{proof}

First of all we fix a number $r_*$ (corresponding to the one denoted
by $r$ in the statement of theorem \ref{main}) determining the order
of normalization we want to reach. In the following we will use the
notation 
$$ 
a\sleq b
$$
to mean: There exists a positive constant $C$ independent of $R$ and
of $N$ (and of all the other relevant parameters, but dependent on
$r_*$, $s$, $\gamma$ and $\alpha$), such that 
$$
a\leq Cb\ .
$$

The proof is based on the
iterative elimination of nonresonant monomials. In order to improve by
one the order of the normalized part of the Hamiltonian we will use a
canonical transformation generated by Lie transform, namely the time 1
flow of a suitable auxiliary Hamiltonian function. So, first of all we
recall some facts about Lie transform, and we introduce some related
tools.

Consider an auxiliary Hamiltonian function $\chi$ and the
corresponding Hamilton equations $\dot z=X_\chi(z)$. Denote by
$T^t$ the corresponding flow and by $\Tr:=T^1\equiv T^t\big|_{t=1}$
the time 1 flow.

\begin{definition}
\label{Lie}
The canonical
transformation $\Tr$ will be called 
the {\it Lie transform} generated by $\chi$.\qed
\end{definition}  

Given an analytic function $g$, consider the transformed
function $ g\circ\Tr $. Using the relation
$$
\frac d{dt}[g\circ T^t]=\poisson\chi g\circ T^t\ ,
$$
it is easy to see that, at least formally, 
one has
\begin{equation}
\label{p.11}
g\circ \Tr=\sum_{l=0}^\infty g_l\ ,
\end{equation}
with $g_l$ defined by
\begin{equation}
\label{p.12}
\quad g_0=g\ ,
g_l=\frac1l\poisson{\chi}{g_{l-1}}\ ,\quad l\geq1\ .
\end{equation}
Then, provided one is able to show the convergence of the series,
(\ref{p.11}) gets a rigorous meaning. 

We will use (\ref{p.11}) to show that, if $g$ and $\chi$ have
$s$--tame modulus then the same is true also for $g\circ\Tr$. To this
end, since infinite sums are involved we have to introduce a suitable
notion of convergence.

\subsection{Properties of the functions with tame modulus}
\label{tmclass}

In this section we will use only the complex variables $\xi,\eta$
defined by (\ref{variables}).  When dealing with such variables, we
will continue to denote by $z$ a phase point (i.e.
$z=(\ldots,\xi_{l},\ldots,\xi_{1},\eta_{1},\ldots,\eta_{l},\ldots)$)
but we will use complex spaces, so in this context $\Ph_s$ will denote
the complexification of the phase space and $B_{s}(R)$ the complex
ball of radius $R$ centered at the origin.

\begin{definition}
\label{normatame3}
Let $X$ be an $s$--tame vector field homogeneous as a polynomial in
$z$; the infimum of the constants $C_s$ such that the inequality 
\begin{eqnarray}
\nonumber
\norma{\tilde X(z^{(1)},...,z^{(r)}) }&\leq& C_s
\frac{1}{r} \sum_{l=1}^{r}
\norma{z^{(1)}}_{1}....\norma{z^{(l-1)}}_{1} \norma{z^{(l)}}_{s}
\norma{z^{(l+1)}}_{1}...\norma{z^{(r)}}_{1}
\\
&&
\forall
z^{(1)},...,z^{(r)}\in\Ph_s 
\label{qas}
\end{eqnarray}
 holds will
be called {\it tame norm $s$ of $X$} (or tame $s$ norm). Such a norm
will be denoted by $\left|X\right|_s^T$.\qed
\end{definition}

\begin{definition}
\label{normatame}
Let $f\in T_M^s$ be a homogeneous polynomial.
The tame norm $s$ of $X_{\mod f}$ will be denoted by $|f|_s$.\qed
\end{definition}

It is useful to introduce a simple notation for the r.h.s. of
(\ref{qas}), so we will write
\begin{equation}
\norma{(z^{(1)},...,z^{(r)})}_{s,1}:=\frac{1}{r} \sum_{l=1}^{r}
\norma{z^{(1)}}_{1}....\norma{z^{(l-1)}}_{1} \norma{z^{(l)}}_{s}
\norma{z^{(l+1)}}_{1}...\norma{z^{(r)}}_{1} 
\label{multinorm}
\end{equation}
Moreover, we will often denote by $w\equiv (z^{(1)},...,z^{(r)})$ a
multivector. Thus 
the quantity (\ref{multinorm}) will be simply denoted by
$\norma{w}_{s,1}$. 

\begin{remark}
\label{normat2}
The tame $s$ norm of a polynomial Hamiltonian $f$ of degree $r+1$ is
given by
\begin{equation}
\label{tamemap1}
|f|_s:=\sup 
\frac{\norma{\tilde X_{\mod f} (w) }_s}  
{\norma{w}_{s,1}
}
\end{equation}
where the sup is taken over all the multivectors 
$$
w=(z^{(1)},...,z^{(r)})
$$
such that $z^{(l)}\not=0$ for any $l$, and $\norma{w}_{s,1}$ is
defined by (\ref{multinorm}).\qed
\end{remark}

\begin{remark}
\label{sup}
Since all the components of the multilinear form 
$\widetilde{ \mod{X_{f}}}$ are positive the above supremum can be
taken only on the positive `octant' on which all the components of
each of the vectors $z^{(l)}$ are positive.\qed
\end{remark}

\begin{remark}
\label{bana1}
If $f\in T^s_M$ is a homogeneous polynomial of degree $r+1$ then one has
\begin{equation}
\label{1.1}
\|X_{f}(z)\|_s\leq \norma{X_{\mod f}(z)}_s\leq |f|_s\norma z_{1}^{r-1}
\norma z_s
\end{equation}\qed
\end{remark}

\begin{definition}
\label{normatame2}
Let $f\in T_M^s$ be a non homogeneous polynomial. Consider its Taylor
expansion 
$$
f=\sum f_r
$$
where $f_r$ is homogeneous of degree $r$. For $R>0$ we will denote 
\begin{equation}
\label{snormav}
\snorma{f}_{s,R}:=\sum_{r\geq 2}|{f_r}|_s R^{r-1}\ .
\end{equation}
Such a definition extends naturally to the set of analytic functions
such that (\ref{snormav}) is finite. 
The set of the functions of class $T_M^s$ with a finite
$\snorma{f}_{s,R}$ norm will be denoted by $T_{s,R}$.\qed  
\end{definition}

\begin{definition}
\label{norma1}
Let $f$ be an analytic function whose vector field is analytic as a
map from $B_s(R)$ to $\Ph_s$.
We denote
$$
\norma{X_f}_{s,R}:=\sup_{\norma z_s\leq R}\norma{X_f(z)}_s
$$
\qed
\end{definition}

\begin{remark}
\label{rem2}
With the above definitions, for any $f\in T_{s,R}$,
one has 
\begin{equation}
\label{n.e}
\norma{X_{f}}_{s,R}\leq \norma{X_{\mod f}}_{s,R} \leq
\snorma{f}_{s,R}
\end{equation}\qed
\end{remark}

\begin{remark}
\label{banach}
The norm $\snorma f_{s,R}$ makes the space $T_{s,R}$ a Banach space.\qed
\end{remark}

A key property for the proof of theorem \ref{main} is related to the
behaviour of functions of class $T_M^s$ with respect to the
decomposition of the phase variables into ``variables with small
index'' and ``variables with large index''.  To be precise we fix some
notations. Corresponding to a given $N$ we will denote $\bar z\equiv
(\bar \xi_j,\bar\eta_j)_{j\leq N}=(\xi_j,\eta_j)_{j\leq N}$ for the
first $N$ canonical variables and $\hat z\equiv
(\hat\xi_j,\hat\eta_j)_{j>N}=(\xi_j,\eta_j)_{j>N}$ for the remaining
ones.

We have the the following important

\begin{lemma}
\label{Ngrande}
Fix $N$ and consider the decomposition $z=\bar z+\hat z$, as above.
Let $f\in T_M^s$ be a polynomial of degree less or equal than
$r_*+2$. Assume that $f$ has a zero of order three in the variables
$\hat z$, then one has
\begin{equation}
\label{sti.N}
\norma{X_{f}}_{s,R}\sleq \frac{\snorma{f}_{s,R}}{N^{s-1}}\
. 
\end{equation}
\end{lemma}

The proof of this lemma is based on two facts: (i) if a sequence
$z\in\Ph_s$ has
only components with large index (i.e. larger than $N$), then its
$\Ph_{1}$ norm is bounded by its $\Ph_s$ norm divided by
$N^{s-1}$; (ii) according to the tame property, in the estimate
of $\norma{X_f}_{s,R}$ the quantity $\norma{\hat z}_{1}$ appears at
least once. The actual prove is slightly complicated due to the
different behaviour of the different components of the Hamiltonian
vector field with respect to the variables with small and large
index. For this reason it is deferred to the appendix A.

\begin{lemma}
\label{poisson}
Let $f,g\in T_M^s$ be homogeneous polynomials of degrees $n+1$ and
$m+1$ respectively, then one has $\left\{f;g\right\}\in T_M^s$ with 
\begin{equation}
\label{sti1}
\left|\left\{f;g\right\}\right|_{s}\leq (n+m)|f|_s|g|_s 
\end{equation}
\end{lemma} 

The proof is based on the following three facts: (i) the vector field of the
Poisson brackets of two functions is the commutator of the vector
fields of the original functions; (ii) in case of polynomials 
the commutator can be computed in terms of the composition of the
multilinear functions associated to the original functions; (iii) the
composition of two tame multilinear functions is still a tame
multilinear function. However the proof requires some attention due to
the moduli and the symmetrization required in the definition of the
class $T_M^s$. For this reason it is deferred to the appendix \ref{tech}.

\begin{lemma}
\label{poi1}
Let  $h,g\in T_{s,R}$, then for any positive 
$d<R$ one has $\left\{f;g\right\}\in T_{s,R-d}$ and
\begin{equation}
\label{poi}
\snorma{\{h;g\}}_{s,R-d}\leq\frac1{d}\snorma h_{s,R}\snorma
g_{s,R} \ .
\end{equation}
\end{lemma}

\proof Write $h=\sum_{j}h_{j}$ and $g=\sum_{k}g_{k}$ 
with $h_{j}$ homogeneous of degree $j$ and similarly for $g$, we
have 
$$
\{h;g\}=\sum_{j,k}\{{h_{j}};{g_{k}}\}\ .
$$
Now each term of the series is estimated by
\begin{eqnarray*}
\snorma{\poisson{h_{j}}{g_k}}_{s,R-d}=\left|
\poisson{h_{j}}{g_{k}}\right|_s (R-d)^{j+k-3}
\\
\leq
\left|h_j\right|_s|g_k|_s (j+k-2) (R-d)^{j+k-3}
\\
\leq 
 \left|h_j\right|_s |g_k|_s \frac1{d}
R^{j+k-2}  = 
\frac1{d}\snorma{h_j}_{s,R}\snorma{g_k}_{s,R}\ ,
\end{eqnarray*}
where we used the inequality 
\begin{equation}
\label{rm}
k(R-d)^{k-1}<\frac{R^k}{d}\ ,
\end{equation}
which holds for any positive $R$ and $0<d<R$. Then the thesis
follows. 
\qed

We estimate now the terms of the series (\ref{p.11},\ref{p.12}) 
defining the  Lie transform. 

\begin{lemma}
\label{t.5}
Let $g\in T_{s,R}$ and $\chi\in T_{s,R}$ be two analytic 
functions; denote by $g_n$ the functions defined recursively by
(\ref{p.12}); 
then, for any positive $d<R$, one has 
$g_n\in T_{s,R-d}$, and the following estimate holds
\begin{equation}
\label{t.23}
\snorma{g_n}_{s,R-d}\leq \snorma g_{s,R}
\left(\frac{e}{d}\snorma\chi _{s,R} \right)^n \ .
\end{equation}
\end{lemma}

\proof Fix $n$, and denote 
$ \delta:=d/n$, we look for a sequence $C^{(n)}_l$ such that 
$$
\snorma{g_l}_{s,R-\delta l}\leq C^{(n)}_l\ ,\quad\forall l\leq n\ .
$$
By \eqref{poi} this sequence can be defined by
$$
C^{(n)}_0=\snorma g_{s,R}\ ,\quad 
C^{(n)}_l=
\frac{1}{l\delta }C^{(n)}_{l-1}
\snorma{\chi}_{s,R}= 
\frac{n}{ld}
C^{(n)}_{l-1}\snorma{\chi}_{s,R}\ .
$$
So one has
$$
C^{(n)}_n
=\frac1{n!}\left(\frac{n
\snorma\chi_{s, R}}{d}\right)^n\snorma 
g_{s,R}\ .
$$
Using the inequality 
$
n^n<n!e^n,
$
which is easily verified by writing the iterative definition of
$n^n/n!$, one has the thesis. 
\qed

\begin{remark}
\label{l.d}
Let $\chi$ be an analytic function with Hamiltonian vector
field which is analytic as a map from $B_s(R)$ to $\Ph_s$, fix $d<R$. 
Assume $\norma{X_\chi}_{s,R}<d$ and consider the 
time $t$ flow $T^t$ of $X_\chi$. Then, 
for $|t|\leq1$, 
one has
\begin{equation}
\label{e.t}
\sup_{\norma{z}_s\leq R-d}\norma{T^t(z)-z}_s\leq
\norma{X_\chi}_{s,R} \ .
\end{equation}\qed
\end{remark}

\begin{lemma}
\label{la}
Consider $\chi$ as above and 
let $g:B_s\left(R\right)\to\C$ be an analytic function with vector
field analytic in $B_s\left(R\right)$, fix $0<d<R$ assume
$\norma{X_\chi}_{s,R}\leq d/3$, then, for $|t|\leq1$, one has
$$
\norma{X_{g\circ T^t}}_{s,R-d }\leq 
\left(1+\frac{3}{d}\norma{X_{\chi}}_{s,R}\right)
\norma{X_g}_{s,R}
$$
\end{lemma}
\noindent {For the proof see \cite{bam99} proof of lemma 8.2}. 

\begin{lemma}
\label{homol}
Let $f$ be a polynomial in $T_M^s$
which is at most quadratic in the variables $\hat
z$. 

There exists $\chi, \Ze\in T_{s,R}$  with $\Ze$ in
$(\gamma,\alpha,N)$--normal form such that
\begin{equation}
\label{homo1}
\poisson{H_0}{\chi}+\Ze=f\ .
\end{equation}
Moreover
$\Ze$ and $\chi$ fulfill the estimates
\begin{equation}
\label{es.homo}
\snorma{\chi}_{s,R}\leq \frac{N^\alpha}{\gamma}\snorma{f}_{s,R}\
,\quad \snorma{\Ze}_{s,R}\leq \snorma{f}_{s,R}
\end{equation}
\end{lemma}
\proof Expanding $f$, in Taylor series, namely
$$
f(\xi,\eta)=\sum_{j,l}{f_{jl}}\xi^j\eta^l
$$ 
and similarly for $\chi$ and $\Ze$,
the
equation (\ref{homo1}) becomes an equation for the coefficients of $f$,
$\chi$ and $\Ze$, namely
$$
\im\omega\cdot (j-l)\chi_{jl}+\Ze_{jl}=f_{jl}
$$ 
We define
\begin{eqnarray}
\label{defZ}
&&\Ze_{jl}:=f_{jl}\ ,\quad j,l\quad  \text{such that}\quad |\omega\cdot
(j-l)| <\frac{\gamma}{N^\alpha}
\\
\label{defchi}
&&\chi_{jl}:=\frac{f_{jl}}{\im\omega\cdot (j-l)}\ ,\quad j,l\quad
\text{such that}\quad |\omega\cdot (j-l)| \geq\frac{\gamma}{N^\alpha}\ ,
\end{eqnarray}
and $\Ze_{jl}=\chi_{jl}=0$ otherwise.
By construction, $\Ze$ and $\chi$ are in $T_M^s$. Further,
since $f$ is at most quadratic in the variables $\hat z$ one has
$\sum_{k>N} (j_k+l_k)\leq 2$ and
thus $\Ze$ is in $(\gamma,\alpha,N)$--normal form.  The estimates (\ref{es.homo}) immediately follow from the
definition of the norm.\qed

\begin{lemma}
\label{t.52}
Let $\chi\in T_{s,R}$ be the solution of the
homological equation (\ref{homo1}) with $f\in T_M^s$. Denote by
$H_{0,n}$ the functions defined recursively as in (\ref{p.12}); for any
positive $d<R$, one has $H_{0,n}\in T_{s,R-d}$, and the following
estimate holds
\begin{equation}
\label{t.23'}
\snorma{H_{0,n}}_{s,R-d}\leq 2\snorma f
_{s,R}
\left(\frac{e}{d}\snorma\chi _{s,R} \right)^n \ .
\end{equation}
\end{lemma}
\proof The idea of the proof is that, using the homological equation one gets
$H_{0,1}= \Ze-f\in T_M^s$. Then proceeding as in the proof of lemma
\ref{t.5} one gets the result.\qed

\subsection{The main lemma and conclusion of the proof}
\label{iterative}

The main step of the proof is a proposition allowing to increase by
one the order of the perturbation. 
As a preliminary step expand $P$ in
Taylor series up to order $r_*+2$:
\begin{equation}
\label{tayex}
P=P^{(1)}+\resto_*\ ,\quad P^{(1)}:=\sum_{l=1}^{r_*}P_l
\end{equation}
where $P_l$ is homogeneous of degree $l+2$ and $\resto_*$ is the remainder
of the Taylor expansion. 

\begin{remark}
\label{rem.3}
From assumption (H) it  follows that there exist 
$R_\sharp$ and $A$ (depending on $r_{*}$ ans $s$) such that one has
\begin{eqnarray}
\label{sti.6}
\snorma{P^{(1)}}_{s,R}\leq A R^2\ ,\quad \forall R\leq R_\sharp 
\\
\label{sti.7}
\norma{X_{\resto_*}}_{s,R}\leq A R^{r_*+2}\ ,\quad \forall R\leq
R_{\sharp}
\end{eqnarray}\qed
\end{remark}

Consider now the analytic Hamiltonian 
\begin{equation}
\label{ham}
H_T=H_0+P^{(1)}
\end{equation}
 and introduce the
complex variables $(\xi,\eta)$ defined by (\ref{variables}). Clearly
the constants $A, R_\sharp$ can be chosen in such a way that
(\ref{sti.6}) holds also in the complex variables.   

From now on we consider $R_\sharp$ as fixed. In
the statement of the forthcoming iterative lemma we will use the
following notations: For any positive $R$, define $\delta:=R/2r_*$ and
$R_r:=R-r\delta$.

\begin{proposition}
\label{po}
{\bf Iterative Lemma. } Consider the
Hamiltonian (\ref{ham}) and let $N$ be a fixed integer. Define $R_*$ 
by 
\begin{equation}
\label{R*}
R_*:=\frac{\gamma}{24e r_* N^\alpha A}
\end{equation}
and assume
\begin{equation}
\label{4.7}
R\leq R_{\sharp}\ ,\quad\frac{R}{R_*}\leq \frac{1}{2}
\end{equation}
then, for any $r\leq r_*$  there exists a canonical
transformation $\Tr^{(r)}$ which puts (\ref{ham}) in the form
\begin{equation}
\label{3.1}
H^{(r)}:=H_T\circ\tk= H_0+ \Ze^{(r)}+f^{(r)}+\resto^{(r)}_N+\resto^{(r)}_T\ ,
\end{equation}
where
\begin{itemize}
\item[1)] the transformation $\Tr^{(r)}$ satisfies 
\begin{equation}
\label{stra}
\sup_{z\in B_s(R_r)}\norma{z-\Tr^{(r)}(z)}\sleq 
{N^{\alpha}}R^{2}
\end{equation}
\item[2)] $\zk r$ is a polynomial of degree at most $r+2$ having a
zero of order $3$ at the origin and is in $(\gamma,\alpha,N)$--normal
form; $f^{(r)}$ is a polynomial of degree less than $r_*+1$ having
a zero of order $r+3$ at the origin. Moreover the following estimates
hold
\begin{eqnarray}
\label{3.2}
\snorma{\zk{r}}_{s,R_r}&\leq&\left\{ 
\begin{matrix}
0 & r=0
\cr
AR^{2}\sum_{l=0}^{r-1}\left(\frac{R}{R_*}\right)^{l} & r\geq1
\end{matrix}
\right.
\\
\label{3.21}
\snorma{\fk r}_{s,R_r}&\leq& A R^{2}\left(\frac{R}{R_*}\right)^{r}
\end{eqnarray}
\item[3)] the remainder terms, $\resto^{(r)}_N$ and $\resto^{(r)}_T$
satisfy for any $s\geq 1$ 
\begin{eqnarray}
\label{RT}
\norma{X_{\resto^{(r)}_T}}_{s,R_r}\sleq R^2\left(\frac{R}{R_*}
\right)^{r_*}
\\
\label{RN}
\norma{X_{\resto^{(r)}_N}}_{s,R_r}\sleq \frac{
R^2}{N^{s-1}}\ .
\end{eqnarray} 

\end{itemize}
\end{proposition}

\proof We proceed by induction. First remark that the theorem is
trivially true when $r=0$ with $\Tr^{(0)}= I$, $\Ze^{(0)}=0$, $\fk 
0 =P^{(1)}$, $\resto^{(0)}_N = 0$ and $\resto^{(r)}_T = 0$.
Then we split $\fk r$ ($P^{(1)}$ in the
case $r=0$) into an effective part and a remainder.  Consider the
Taylor expansion of $\fk{r}$, {\it in the variables $\hat z$
only}. Write
$$
\fk{r}=\fk{r}_0+\fk{r}_N
$$
where $\fk{r}_0$ is the truncation of such a series at second order
(it contains at most terms quadratic in $\hat z$) and $\fk{r}_N$ is the
remainder of the expansion.  Since both $\fk r_{N}$ and $\fk r_0$ are
truncations of $\fk r$, one has
$$
\snorma{\fk{r}_N}_{s,R_{r}}\leq \snorma{\fk{r}}_{s,R_r}\ ,\quad
\snorma{\fk{r}_0}_{s,R_{r}}\leq \snorma{\fk{r}}_{s,R_r} .
$$
Consider the truncated Hamiltonian
\begin{equation}
\label{tru}
H_0+\zk r+\fk r_0\ .
\end{equation}
We look for a Lie transform, $\Tr_r$, eliminating the non normalized part 
of order $r+2$ in the truncated Hamiltonian.
Let $\chi_r$ be the Hamiltonian  generating  $\Tr_r$. Using the
formulae (\ref{p.11},\ref{p.12}) one writes
\begin{eqnarray}
\label{lie.1}
\left(H_0+\zk r+\fk r_0\right)\circ \Tr_r = H_0+\zk{r}
\\
\label{lie.2}
+\poisson{\chi_r}{H_0}+\fk r_0
\\
\label{lie.3}
+\sum_{l=1}^{l_z}\zk r_l+\sum_{l=1}^{l_f}\fk
 r_{0,l}+\sum_{l=2}^{l_f+1}H_{0,l}
\\
\label{lie.4}
+ 
\sum_{l>l_z} \zk r_l+\sum_{l>l_f} \fk
 r_{0,l}+\sum_{l>l_f+1}H_{0,l}
\end{eqnarray}
where
\begin{equation}
\label{elle}
l_z:=\left[\frac{r_*-1}{r}\right]\ ,\quad
l_f:=\left[\frac{r_*}{r}-1\right]\ .
\end{equation}
Then it easy to see that (\ref{lie.1}) is the already normalized part
of the transformed Hamiltonian, (\ref{lie.2}) contains the part of
degree $r+2$, from which all the non normalized terms have to be
eliminated by a suitable choice of $\chi_r$, (\ref{lie.3}) contains
all the terms of degree between $r+3$ and $r_*+2$ and finally
(\ref{lie.4}) contains only terms of order higher than $r_*+2$, which
therefore can be considered as an irrelevant remainder for the rest of
the procedure (actually they will be incorporated in $\resto_T$).

We first use lemma \ref{homol} to determine $\chi_r$ as  the
solution of the equation
\begin{equation}
\label{homo.4}
\poisson{\chi_r}{H_0}+\fk r_0=\Ze_r
\end{equation}
with $\Ze_r$ in normal form. Then, by  \eqref{es.homo} and by (\ref{3.21})
one has the estimates
\begin{equation}
\label{es.homo1}
\snorma{\chi_r}_{s,R_r}\leq \frac{N^\alpha}{\gamma}  \
AR^2\left(\frac{R}{R_*} \right)^r
,\quad \snorma{\Ze_r}_{s,R_r}\leq AR^2\left(\frac{R}{R_*} \right)^r \ .
\end{equation}
In particular, in view of \eqref{e.t}, estimate \eqref{stra} is 
proved at rank $r+1$. 

Define now $\Ze^{(r+1)}:=\Ze^{(r)}+\Ze_r $, $\fk{r+1}:=$(\ref{lie.3}) 
and
$\resto_{(r,T)}:=(\ref{lie.4})$. From \eqref{es.homo1} the estimate 
\eqref{3.2} holds at rank $r+1$. By lemma \ref{t.5}, denoting
$$
\epsilon:=\frac{e}{\delta}\snorma{\chi_r}_{s,R_r},
$$ 
one has using \eqref{4.7}, \eqref{3.2}, \eqref{3.21} and lemma 
\ref{t.52}
\begin{eqnarray*}
\snorma{\fk{r+1}}_{s, R_{r}-\delta} \leq\sum_{l\geq 1} 2
AR^2\epsilon^l +\sum_{l\geq 1}AR^2\epsilon^l
\left(\frac{R}{R_*}\right)^r +\sum_{l\geq 2}2AR^2 \epsilon^{l-1}  
\left(\frac{R}{R_*}\right)^r 
\\
\leq \frac{\epsilon}{1-\epsilon}2AR^2\left[
1+\left(\frac{R}{R_*}\right)^r+2\left(\frac{R}{R_*}\right)^r
\right]< 12\epsilon AR^2\ .
\end{eqnarray*}
Due to the definition of $R_*$ and to the following estimate of
$\epsilon$,
$$
\epsilon\leq
\frac{e}{\delta}AR^2\left(\frac{R}{R_*}\right)^r
\frac{N^\alpha}{\gamma}\ ,
$$
we deduce that 
$$
\snorma{\fk{r+1}}_{s, R_{r}-\delta} \leq AR^2\left(\frac{R}{R_*}\right)^{r+1}
$$
which is the needed estimate of $\fk{r+1}$.

In a similar way one easily checks that 
\begin{equation}
\label{RT2}
\snorma{X_{\resto_{r,T}}}_{s,R_r-\delta}\sleq
AR^2\left(\frac{R}{R_*}\right)^{r_*+2} \ . 
\end{equation}
Define now 
\begin{equation}
\label{RT1}
\resto^{(r+1)}_T:=\resto^{(r)}_T\circ\Tr_r+\resto_{r,T}\ .
\end{equation}
Remark that \eqref{RT} means that there exists a constant $C^T_r$ such that
$$
\norma{\resto^{(r)}_T}_{s,R_r}\leq C^T_rAR^2\left(\frac{R}{R_*}
\right)^{r_*}\ .
$$
By lemma \ref{la} the first term of (\ref{RT1}) is estimated by
$2C_r^T$. By adding the estimate (\ref{RT2}) one gets the existence of
the constant such that (\ref{RT}) holds also for $\resto_T^{(r+1)}$.

 Concerning the terms at least cubic in $\hat z$, define 
\begin{equation}
\label{RN1}
\resto^{(r+1)}_N:=\left(\resto^{(r)}_N+f_N^{(r)}\right)\circ\Tr_r\ .
\end{equation}
Using lemma \ref{Ngrande} one estimates $\norma{X_{f_N^{(r)}}}_{s,R_r}$.
Adding the iterative estimate of $X_{\resto^{(r)}_N}$ and estimating the
effects of $\Tr_r$ by lemma \ref{la} one gets (\ref{RN}) at rank $r+1$.
\qed 

\begin{corollary}
\label{mc}
For any $r_{*}\geq 1$ and $s\geq 1$ there exists a constant $C$ such 
that for any $R$ satisfying
$$
R\leq \frac{C}{N^{\alpha}}
$$
the following holds true: There exists a canonical transformation
$\Tr_r :B_s(R/3)\to B_s(R)$ fulfilling
$$
\norma{z-\Tr_r(z)}_s\sleq (N^{\alpha}R)^2 R \ 
$$ 
such that the transformed Hamiltonian has the form
\begin{equation}
\label{normal}
H^{(r_*)}=H_0+\Ze^{(r_*)}+\resto_N+\resto_{T}+\resto_*\circ\Tr_r
\end{equation}
where $\Ze^{(r_*)}$ is in $(\gamma,\alpha,N)$--normal form, and the
remainders fulfill the following estimate
\begin{eqnarray}
\label{RT4}
\norma{\resto_*\circ\Tr_r}_{s,R/2}\ ,\ \norma{\resto_T}_{s,R/2}\sleq
R^2(RN^\alpha)^{r_*} \\
\label{RN4}
\norma{\resto_N}_{s,R/2}\sleq \frac{R^{2}}{N^{s-1}}
\end{eqnarray}
\end{corollary}

\noindent {\it End of the proof of theorem \ref{main}.} 
 To conclude the proof we choose $N$ and $s$ in order to obtain
that both $\resto_{T}$ and $\resto_N$ are small. First take $N=R^{-a}$
with a still undetermined $a$. Then in order to obtain that $\resto_T$
is of order larger than $R^{r_*+1}$ one has to choose $a$ so that
$a<(1/r_*\alpha)$. We choose $a:=(1/2r_*\alpha)$. Inserting in the
estimate of $\resto_N$ one has that this is surely smaller than
$R^{r_*+1}$ if $s>2\alpha r_*^2+1$. With this choice the theorem is
proved. \qed

\section{Verification of the Nonresonance Properties}
\label{B}

\subsection{Dirichlet frequencies of the wave equation}
\label{v.400}

The proof follows very closely the proof of theorem 6.5 
of \cite{Bam03}. We repeat the main steps for completeness and because
we will use a variant of it  in the next subsection. We use the 
notations introduced in section \ref{dirichlet}.
We fix $V_0$ and $r$ once for all and denote by $C$ any 
 constant depending only on $V_{0}$ and $r$.

\begin{lemma}\label{det}
{For any $K\leq N$, consider $K$ indexes $\bji 1<...<\bji K\leq N$;
consider the determinant
\begin{equation}
D:=\left|
\begin{matrix}
\omega_{\bji 1}& \omega_{\bji 2}&.& .&.&\omega_{\bji K} 
\\
\der \null {\bji 1} & \der \null {\bji 2} & .& .&.&\der \null {\bji K}
\\
.& .& .& .& .&.
\\
.& .& .& .& .&.
\\
\der{K-1}{\bji 1}& \der{K-1}{\bji 2}& .& .&.&\der {K-1}{\bji K}
\end{matrix}
\right|
\end{equation}
One has
\begin{eqnarray}
D&=&\pm \left[\prod_{j=1}^{K-1}\frac{(2j-3)!}{2^{j-2}(j-2)!2^j}
  \right]\left(\prod_{l}\omega_{i_l}^{-2K+1} \right) 
  \left(\prod_{1\leq l<k\leq
  K} (\lambda_{\bji l}-\lambda_{\bji k} )\right)
\\
\label{v.401}
&\geq& 
\frac{C}{N^{2K^2}}
. \end{eqnarray}
}\end{lemma}

\proof First remark that, by explicit computation one has 
\begin{equation}
\frac{d^j\omega_i}{dm^j}=\frac{(2j-3)!}{2^{j-2}(j-2)!2^j}
\frac{(-)^{j+1}} 
{(\lambda_{i}+m)^{j-\frac12}}\ .\label{df}
\end{equation}
Substituting (\ref{df}) in the l.h.s. of (\ref{det}) we get the
determinant to be computed. Factorize from the $l-th$ column the term
$(\lambda_{\bji l}+m)^{1/2}$, and from the $j-th$ row the term
$\frac{(2j-3)!}{2^{j-2}(j-2)!2^j}$. The determinant becomes, up to the
sign,
\begin{eqnarray}
\left[\prod_{l=1}^{K}\omega_{\bji l}\right]
\left[\prod_{j=1}^{K-1}\frac{(2j-3)!}{2^{j-2}(j-2)!2^j} \right]  \nonumber
\\
\times
\left|
\begin{matrix}
1& 1& 1 &. & . & . & 1
\cr
x_{\bji 1}& x_{\bji 2}& x_{\bji 3}&.&.&.&x_{\bji K}
\cr
x_{\bji 1}^2& x_{\bji 2}^2& x_{\bji 3}^2&.&.&.&x_{\bji K}^2
\cr
.& .& .& .& .&.&.
\cr
.& .& .& .& .&.&.
\cr
.& .& .& .& .&.&.
\cr
x_{\bji 1}^{K-1}& x_{\bji 2}^{K-1}& x_{\bji 3}^{K-1}&.&.&.&x_{\bji K}^{K-1}
\end{matrix}
\right|
\label{van1}
\end{eqnarray}
where we denoted by $x_{j}:=(\lambda_{j}+m)^{-1}\equiv\omega_{j}^{-2}$.
The last determinant is a Vandermond determinant whose value is given
by
\begin{equation}
\prod_{1\leq l<k\leq K}(x_{\bji l}-x_{\bji k})=\prod_{1\leq l<k\leq K}
\frac{\lambda_{\bji
    k}-\lambda_{\bji l}}{\omega_{\bji l}^2\omega_{\bji k}^2}=
\left(\prod_{1\leq l<k\leq K}(\lambda_{\bji l}- \lambda_{\bji
  k})\right)\prod_{l=1}^K 
    \omega_{j_l}^{-2K} 
\label{van}\ .
\end{equation}
Using the asymptotic of the frequencies and the fact that all the
eigenvalues are different one gets the thesis also the second of
(\ref{v.401}).\qed 

From \cite{ben85} appendix B we learn
\begin{lemma}\label{m1.1}
 Let
$u^{(1)},...,u^{(K)}$ be $K$ independent vectors with
$\norma{u^{(i)}}_{\ell^1}\leq1$. Let $w\in\R^K$ be an arbitrary
vector, then there exist $i\in[1,...,K]$, such that
$$
|u^{(i)}\cdot w|\geq\frac{\norma w_{\ell^1}\det(u^{(1)},\ldots,u^{(K)})}
{K^{3/2}}\ .
$$
\end{lemma}

Combining Lemmas \ref{det} and \ref{m1.1} we deduce

\begin{corollary}
\label{m1.2}Let $w\in\R^{\infty}$ be
a vector with $K$ components different from zero, namely those
with index $\bii1,...,\bii{K}$; assume $K\leq N$, and 
$\bii1<...<\bii{K}\leq N$.
Then, for any $m\in [m_0,\Delta]$ there exists an index $i\in[0,...,K-1]$ such
that
\begin{equation}
\left|w\cdot\frac{d^i\omega}{dm^i}(m)\right|\geq
C\frac{\norma{w}_{\ell^1}}{N^{2K^2+2}}
\end{equation}
where $\omega$ is the frequency vector.
\end{corollary}

Now we need the following lemma from \cite{you97a}.

\begin{lemma}
\label{v.112}
(Lemma 2.1 of \cite{you97a}) Suppose that $g(\tau)$ is
$m$ times differentiable on an interval $J\subset\R$. Let
$J_h:=\left\{\tau\in J\ :\ |g(\tau)|<h\right\}$, $h>0$. If on $J$,
$\left|g^{(m)}(\tau)\right|\geq d>0$, then $|J_h|\leq M h^{1/m}$,
where 
$$
M:=2(2+3+...+m+d^{-1})\ .
$$
\end{lemma}

For any $k\in \Z^N$ with $\vak\leq r$ and for any $n\in \Z$, define 
\begin{equation}
\label{rt}
\resto_{kn}(\gamma,\alpha):=\left\{m\in[m_0,\Delta]\ :\
\left|\sum_{j=1}^N k_j\omega_j+n\right|<\frac{\gamma}{N^{\alpha}}\right\}
\end{equation}

Applying lemma \ref{v.112} to the function
$\sum_{j=1}^{N}k_j\omega_j+n$ and using corollary \ref{m1.2} we get 

\begin{corollary}
\label{m.2}
Assume $|k|+|n|\not=0$, then
\begin{equation}
\left|\resto_{kn}(\gamma,\alpha )\right|\leq C(\Delta-m_0)
\frac{\gamma^{1/r}}{N^{\varsigma}}\label{m.21}
\end{equation}  
with  $\varsigma=\frac{\alpha}{r}-2r^2-2$. 
\end{corollary}

\begin{lemma}
\label{v.001}
Fix $\alpha>2r^3+r^{2}+5r$. For any positive $\gamma$ small enough there
exists a set $\I_\gamma\subset [m_0,\Delta]$ such that $\forall
m\in\I_\gamma$ one has that for any $N\geq 1$ 
\begin{equation}
\label{v.002}
\left|\sum_{j=1}^N k_j\omega_j+n\right|\geq\frac{\gamma}{N^{\alpha}}
\end{equation}
for all $k\in\Z^{N}$ with $0\not=|k|\leq r$ and for all
$n\in\Z$. Moreover,
\begin{equation}
\label{v.003}
\left|[m_0,\Delta]-\I_\gamma\right|\leq C \gamma^{1/r}\ .
\end{equation} 
\end{lemma}
\proof
Define $\I_\gamma:=\bigcup_{nk}\resto_{nk}(\gamma,\alpha)$. Remark that,
from the asymptotic of the frequencies, the argument of the modulus
in (\ref{v.002}) can be small only if $|n|\leq C rN$, By \eqref{m.21} one
has
$$
\left|\bigcup_{k}\resto_{nk}(\gamma,\alpha)\right|\leq
\sum_{k}\left|\resto_k(\gamma,\alpha)
\right|<C\frac{N^{r}(\Delta-m_0)\gamma^{1/r}}{N^{\varsigma}} \ ,
$$ summing over $n$ one gets an extra factor $rN$.  Provided $\alpha$
is chosen according to the statement, one has that the union over $N$
is also bounded and therefore the thesis holds.\qed

Denote $\omega^{(N)}:=(\omega_1,...,\omega_N)$, then the
statement of Theorem \ref{nres} is equivalent to

\begin{lemma}
\label{v.004}
For any $\gamma$ positive and small enough, there exist
a set $\J_\gamma$  satisfying, $\va{[m_0,\Delta]-\J_\gamma}\to 0$ when 
$\gamma \to 0$,
and a real number $\alpha'$ such that for any $m\in\J_{\gamma}$ one 
has for $N\geq 1$
\begin{equation}
\label{v.005}
\left|\omega^{(N)}\cdot
k+\epsilon_1\omega_j+\epsilon_2\omega_l\right|\geq \frac{\gamma}{N^{\alpha'}}
\end{equation}
for any $k\in\Z^N$, $\epsilon_i=0,\pm1$, $j\geq l>N$, and
$|k|+|\epsilon_1|+|\epsilon_2|\not=0$. 
\end{lemma}

\proof
The case $\epsilon_1=\epsilon_2=0$ reduces to the previous lemma with
$n=0$. Consider the case $\epsilon_1=\pm1$ and $\epsilon_2=0$. In view 
of the
asymptotic of the frequencies, the argument of the modulus can be
small only if $j<2rN$. Thus to obtain the result one can simply apply
lemma \ref{v.001} with $N':=2rN$ in place of $N$ and $r':=r+2$ in place 
of $r$. This just amounts to
a redefinition of the constant $C$ in (\ref{v.003}). The argument is
identical in the case $\epsilon_1\epsilon_2=1$.

Consider now the case $\epsilon_1\epsilon_2=-1$. Here the main remark
is that 
\begin{equation}
\label{v.007}
\omega_j-\omega_l=j-l+a_{jk}\ \text{with} \
\left|a_{jl}\right|\leq\frac{C}{l} 
\end{equation}
So the quantity to be estimated reduces to
$$
\omega^{(N)}\cdot
k\pm n\pm a_{jl}\ ,\quad n:=j-l
$$ If $l>2CN^\alpha/\gamma$ then the $a_{jl}$ term represent an irrelevant
correction and therefore the lemma follows from lemma \ref{v.001}. In
the case $l\leq 2CN^\alpha/\gamma$ one reapplies the same lemma with
$N':= 2CN^\alpha/\gamma$ in place of $N$ and $r':=r+2$ in place of $r$.
As a consequence one has
that the nonresonance condition (\ref{v.005}) holds provided
$\alpha'=\alpha^2\sim r^6$ and assuming that $m$ is 
in a set whose complement has its measure
estimated by a constant times $\gamma^\frac{\alpha+1}r$.\qed

\subsection{Periodic frequencies of the wave equation}
\label{periodic wave}

We use the notations of section \ref{periodic}.  We fix $r$ and 
$\sigma$ once for all and 
denote by $C$ any 
 constant depending only on  $r$ and $\sigma$. We introduce 
$\tau_{0}=\omega_{0}$ and for 
$j\geq 1$,
$$
\gamma_j:=\omega_j-\omega_{-j}\
,\quad \tau_j=\frac{\omega_j+\omega_{-j}}{2} \ .
$$ 
We recall that from the Sturm Liouville theory (cf. \cite{KaMi01}) there exists an
absolute constant $C_\sigma$ such that, for $V\in \V_{R}$,
\begin{equation}
\label{v.701}
\left|\lambda_j-\lambda_{-j}\right|\leq C_{\sigma} Re^{-2\sigma j} \ .
\end{equation}

We are going to prove the following result:
\begin{theorem}
\label{pv.1}
Fix a positive $r\geq 2$, define $\alpha:=200 r^3$. There exists a
positive $R$ and a positive $b$ with the following property: for any
$\gamma>0$ small enough there exists a set $\Sc_\gamma\subset \V_{R}$
and a constant $\beta$
such that
\begin{itemize}
\item[i)] For all $ V\in\Sc_\gamma$ one has that for all $N\geq 1$ the
  following inequality, with $J:={b} \ln\frac N{\gamma^{\beta}}$,  holds
\begin{equation}
\label{pv.2}
\left|\sum_{j=1}^{N}\tau_jk_j+\sum_{j=1}^{J}\gamma_j n_j+
\sum_{j=J+1}^{N}\gamma_jl_j
\right|>\frac{\gamma}{N^\alpha}
\end{equation}
for any $k\in\Z^{N+1}$, $l\in\Z^{N-J}$, $n\in\Z^J$ fulfilling
$|k|+|n|\not=0$, $|k|+|l|+|n|\leq r$. 
\item[ii)] $|\V_{R}-\Sc_\gamma|\leq C\gamma^{(1/4r)}$.
\end{itemize}
\end{theorem}
Using arguments similar to those in the proof of lemma \ref{v.004}, one 
easily concludes that theorem \ref{pv.1} implies theorem \ref{teodiofper}.

The strategy of the proof of theorem \ref{pv.1} is as follow: when 
$n=0$ (and then $k\neq 0$) we only move  the mass $m$ in 
order to move the $\tau_{j}$ in such a way  that the l.h.s. of 
\eqref{pv.2} is not too small. We also use that, for $V\in 
\V_{R}$, $\gamma_{j}$ is exponentially small with respect to $j$ and 
thus the third term in the l.h.s. of 
\eqref{pv.2} is not relevant.
In a second step we consider the case where $n\neq 0$ and we use the 
Fourier coefficients $v_{i}$ to move the $\gamma_{j}$ for small indexes 
$j$.

Fix $k\in\Z^{N}$, $l\in\Z^{N-J}$, $n\in\Z^J$. For any $\alpha\in\R$
and $\gamma>0$ define
\begin{equation}
\label{Sc}
\Sc_{knl}(\gamma,\alpha):=\left\{(m,v_k)\in\V_R\ :\
\left|\sum_{j=1}^{N}\tau_jk_j+\sum_{j=1}^{J}\gamma_j n_j +
\sum_{j=J+1}^{N}\gamma_jl_j
 \right|\leq \frac{\gamma}{N^\alpha} \right\}
\end{equation}

\begin{lemma}
\label{per1}
For $0\neq \vak \leq r$
$$
\left|\Sc_{k00}(\gamma_1,\alpha_1)\right|<C
\frac{\gamma_1^{1/r}}{N^{\varsigma}}
$$
with  $\varsigma=\frac{\alpha_1}{r}-2r^2-2$.
\end{lemma}
\proof In this case the l.h.s. of (\ref{pv.2}) reduces to
$\sum_{j=1}^{N}\tau_jk_j$ and we use $m$ to
move it away from zero. To this end we follow closely the argument of
subsection \ref{v.400}. So, consider the determinant 
\begin{equation}
\label{determinante}
\left|
\begin{matrix}
\tau_{j_1}&.&.&.& \tau_{j_K}
\\ 
\frac{d\tau_{j_1}}{dm }&.&.&.& \frac{d\tau_{j_K}}{dm }
\\
.&.&.&.&.
\\
.&.&.&.&.
\\
.&.&.&.&.
\\
\frac{d^{K-1}\tau_{j_1}}{dm^{K-1} }&.&.&.&
\frac{d^{K-1}\tau_{j_K}}{dm^{K-1} } 
\end{matrix}
\right| \ .
\end{equation}
Using the definition of $\tau$ one immediately has that such a
determinant is the sum of some determinant of the form
(\ref{det}). The key point is that all these determinants have the
same sign and therefore the modulus of the determinant
(\ref{determinante}) is estimated from below by  a negative power
of $N$ (as the determinant (\ref{det})). To see that all the
determinants composing (\ref{determinante}) have the same sign remark
that the sign of (\ref{det}) is determined by
\begin{equation}
\label{det1}
\prod_{1\leq l<k\leq
  K} (\lambda_{\bji l}-\lambda_{\bji k} )
\end{equation}
Choose now arbitrarily two of the determinants composing
(\ref{determinante}). The corresponding products (\ref{det1}) differ
only because they involve indexes with the same modulus, but different
sign. Provided the potential (i.e. $R$) is small enough one has that
the sign of $\lambda_{\pm j}-\lambda_{\pm k}$ does not depend on the
choice of the $\pm$'s. Thus the result on (\ref{determinante})
follows. 

Then the thesis of lemma  follows from the
procedure of the previous section in the same way that lemma \ref{det}  
leads to corollary \ref{m.2} .\qed

\begin{corollary}
\label{per2} For each $\gamma_1 >0$, $\alpha_1 >0$, $R>0$, $r$, $N$
define
\begin{equation}
\label{v.700}
J:=\frac{\alpha_1}{2\sigma}\ln
\left[\left(\frac{rRC_{\sigma}}{\gamma_1}\right)^{1/\alpha_1} N \right]
\end{equation}
then for any $0\neq \vak
\leq r$ and any $l\in\Z^{N-J}$
$$
\left|\Sc_{k0l}(\gamma_1 /2,\alpha_1)\right|<C
\frac{\gamma_1^{1/r}}{N^{\varsigma}}
$$
with  $\varsigma=\frac{\alpha_1}{r}-2r^2-2$.
\end{corollary}
\proof From (\ref{v.701}) one has
\begin{equation}\label{gam}
\left|\gamma_j\right|\leq \frac{1}{2j}C_{\sigma}Re^{-2\sigma j} 
\end{equation}
which, using the definition of $J$ leads to
$$
\left|\sum_{j=J+1}^{N}\gamma_jl_j\right|\leq \frac{C_{\V}R}{2J}e^{-2\sigma
  J}r<\frac{\gamma}{2N^{\alpha_1}} 
$$
Then for
$(m,v_k)\in\Sc_{k00}(\gamma_1,\alpha_1)$,
$$
\left|\sum_{j=1}^{N}\tau_jk_j+\sum_{j=J+1}^{N}\gamma_jl_j
 \right|>\frac{\gamma_1}{N^{\alpha_1}}
 -\frac{\gamma_1}{2N^{\alpha_1}}=\frac{\gamma_1}{2N^{\alpha_1}}\ . 
$$ 
\qed

In the case where $n\neq 0$, we need some informations on the periodic
spectrum of a Sturm Liouville operator. As remarked in section 
\ref{periodic}, in the case
of even potentials the periodic spectrum
is the union of the Dirichlet and the Neumann
spectrum. Thus the forthcoming lemma \ref{d2} has the following
\begin{corollary}
\label{freper}
For $V\in \V_{R}$ one has
\begin{equation}
\label{taugamma}
\frac{\partial \gamma_{j}}{\partial v_{2i}}=-\frac 1 4 
\frac{\tau_{j}}{\sqrt{
  \omega_j\omega_{-j}}}\delta_{ji}+ \resto_{ji} \ , \quad 
  \frac{\partial \tau_{j}}{\partial
  v_{2i}}= -\frac 1 8 
\frac{\gamma_{j}}{\sqrt{
  \omega_j\omega_{-j}}}\delta_{ji} +\resto'_{ji}
\end{equation}
with
\begin{equation}
\label{taugamma1}
|\resto_{ji}| \ ,\ |\resto'_{ji}|\leq   C 
\frac{Re^{-\sigma|j-i|}}{j}\ .
\end{equation}
\end{corollary}

\begin{lemma}
\label{per3}With the definition of $J$ given by (\ref{v.700}), 
there exists a positive $R$ (independent of $J,\gamma,\alpha$) such
that for any $V\in \V_{R}$, if $n\neq 0$, $\vak +\va{n} +\val \leq r$
then
$$
\left|\Sc_{knl}(\gamma_2,\alpha_2)\right|<C\frac{\gamma_2}{\gamma_1}
\frac{1}{N^{\alpha_2-32\alpha_1}}  
$$
\end{lemma}
\proof Let $1\leq i\leq J$ such that $n_i\not=0$ and consider
\begin{equation}
\label{1per}
\frac{\partial}{\partial v_{2i}}
\left(\sum_{j=1}^{N}\tau_jk_j+\sum_{j=J+1}^{N}\gamma_jl_j+ 
\sum_{j=1}^{J}n_j\gamma_j  \right)\ .
\end{equation}
By corollary \ref{freper} this quantity is given by
\begin{equation}
\label{2per}
\sum_{j=1}^{N}\resto_{ji}' k_j+\sum_{j=J+1}^{N}\resto_{ji} l_j+ 
\sum_{j=1}^{J}n_j\resto_{ji}-\frac{n_i}{4}\frac{\tau_{i}}{\sqrt{
  \omega_i\omega_{-i}}}-\frac{k_i}{8}\frac{\gamma_{i}}{\sqrt{
  \omega_i\omega_{-i}}} \ .  
\end{equation}
From (\ref{taugamma1}) one has (noticing that $\frac{e^{-\sigma 
\va{i-j}}}{j}\leq \frac C i$ for all $i,j\geq 1$) 
$$
\left|\resto_{ji}'\right|\ ,\ \left|\resto_{ji}\right| <\frac{CR}{i}\ .
$$ 
Combining this estimate  with \eqref{gam} one gets that the modulus
(\ref{2per}) is estimated from below by
$$
\frac{1}{i}(\frac{n_i} 8 -CRr)>\frac{1}{16J}
$$
taking $R=1/16Cr$. It follows that excising from the domain
of $v_{2i}$ a segment of length $\gamma_2 16J/N^{\alpha_2}$, whose 
normalized measure is
estimated by 
$$
\frac{\gamma_216J}{N^{\alpha_2}}\frac{e^{2J\sigma}}{R}\leq
\frac{16\gamma_2}{N^{\alpha_2}
  R}e^{4J\sigma}=C\frac{\gamma_2}{\gamma_1^{2}}\frac{1}
     {N^{\alpha_2-2\alpha_1}}  
$$ 
one fulfills the nonresonance condition.  \qed

\medskip

\noindent 
{\it Proof of theorem \ref{pv.1}} Fix $\alpha$ as in the
statement and a positive $\gamma$. Define $\alpha_1:=\alpha/4$,
$\alpha_2:=\alpha$, $\gamma_1:=\gamma^{1/4}$,
$\gamma_2:=\gamma$. Take $J$ as in (\ref{v.700}). 
Define
\begin{eqnarray}
\Sc_{\gamma,N}:=\bigcup_{knl}\Sc_{knl}(\gamma,\alpha)
\\
=\left[\bigcup_{n\not=0}\Sc_{knl}(\gamma,\alpha)
  \right]\bigcup 
\left[\bigcup_{k\not=0}\Sc_{k0l}(\gamma,\alpha)
  \right]
\\
\subset 
\left[\bigcup_{n\not=0}\Sc_{knl}(\gamma_2,\alpha_2)
  \right]\bigcup 
\left[\bigcup_{k\not=0}\Sc_{k0l}(\gamma_1,\alpha_1)
  \right]
\end{eqnarray}
thus one can use corollary \ref{per2} and lemma \ref{per3} to get the
estimate
\begin{equation}
\label{v.702}
\left|\Sc_{\gamma,N} \right|\leq C N^{r} \left(
 \frac{\gamma_1^{1/r}}{N^{\varsigma}}+
\frac{\gamma_2}{\gamma_1^2}
\frac{1}{N^{\alpha_2-2\alpha_1}}  
 \right)
\end{equation}
Inserting the definitions of the various parameters one has that both
$\alpha_2-2\alpha_1$ and $\varsigma$ are bigger than $r+2$ and thus
one can define
$$
\Sc_\gamma:=\bigcup_{N}\Sc_{\gamma,N}
$$
and estimate its measure by the sum over $N$ of the r.h.s. of
(\ref{v.702}) getting a convergent series and the result.\qed

\subsection{Frequencies of the 1-d NLS}

Denote by $\lambda_j$ the Dirichlet eigenvalues of $-\partial_{xx}+V$.

\begin{lemma}
\label{d2}
For any $j$ and $k$ and any $V\in\V_R$, with $R$ small enough, one has
\begin{equation}
\label{d.21}
\frac{\partial  \lambda_j}{\partial v_k}(V)
=-\delta_{j,2k} \frac{1}{2}+  \resto_{jk}
\end{equation} 
with 
$$
|\resto_{jk}|\leq C Re^{-\sigma||j|-2k|}\ .
$$
For the Neumann eigenvalues the formula (\ref{d.21}) without the minus
sign  holds.
Moreover, for any
 $j>k,l$ 
\begin{equation}
\label{d22}
\left.\frac{\partial ^2 \lambda_j}{\partial v_k\partial v_l}\right|
_{V=0} =-\delta_{kl} \frac{1}{2}\frac{1}{(2j)^2-k^2}=-\delta_{kl}
\frac{1}{8} \sum_{s\geq1} \frac{(k^2)^{s-1}}{2^{s-1}j^{2s}}
\ .
\end{equation}
\end{lemma} 
\proof Denote $L_0:=-\partial_{xx}+V$ with Dirichlet boundary
conditions (all what follows holds also for the case of Neumann
boundary conditions). Fix $j$ and let $\lambda=\lambda(V)$ be the
$j$--th eigenvalue of $L_0$. As the Dirichlet spectrum is simple, the 
function
$V\mapsto \lambda (V)$ is smooth  and admit a Taylor expansion at any 
order. In particular we are interested in computing
such a Taylor expansion up to second order, i.e. in computing
$d\lambda$ and $d^2\lambda$. The idea is to fix a potential $h$ and to
construct iteratively, by Lyapunof--Schmidt method, the expansion of
$\lambda(V+\epsilon h)$. Thus
one will get
$$
\lambda(V+\epsilon h)=\lambda(V)+\epsilon
d\lambda(V)h+\frac{\epsilon^2}{2} \langle d^2\lambda(V);h,h\rangle+...
$$
which obviously allows to compute $d\lambda$, but also $d^2\lambda$
which is the bilinear form associated to the quadratic form $\langle
d^2\lambda(V);h,h\rangle$. 

So consider the eigenvalue equation 
\begin{equation}
\label{v.500}
(L_0+\epsilon h)\varphi=\lambda\varphi
\end{equation}
and formally expand $\phiv$ and $\lambda$ in power series:
$$
\phiv=\sum_{l\geq 0}\epsilon^l\phiv_l\ ,\quad \lambda=\sum_{l\geq0}
\epsilon^l\lambda_l \ ,
$$
inserting in (\ref{v.500}) and equating terms of equal order one gets
\begin{eqnarray}
\label{v.501}
L_0\phiv_0&=&\lambda_0\phiv_0
\\
\label{v.502}
(L_0-\lambda_0)\phiv_l&=&\psi_l+\lambda_l\phiv_0
\end{eqnarray}
where
$$
\psi_l:=-h\phiv_{l-1}+\sum_{l_0=1}^{l}\lambda_{l_0}\phiv_{l-l_0}\ .
$$ Decompose $L^2=$span$\phiv_0\oplus $(span$\phiv_0)^{\perp}
$=Ker$(L_0-\lambda_0)\oplus $Range $(L_0-\lambda_0)$, and let $P$ be
the projector on the orthogonal to span$\phiv_0$. Taking $\phiv_0$
normalized in $L^2$, eq. (\ref{v.502}) turns out to be equivalent to
\begin{eqnarray}
\lambda_l:=-\langle\psi_l ;\phiv_0\rangle_{L^2}=\langle\phiv_{l-1};
\phiv_0 \rangle_{L^2} 
\\
\phiv_l:=(L_0-\lambda_0)^{-1}P\psi_l\ .
\end{eqnarray} 
By taking $V\in\V_R$, $h=\cos(kx)$ and using (S1) one gets
(\ref{d.21}). By taking $V=0$ and $h=\mu_1\cos(kx)+\mu_2\cos(lx)$ (and
developing the computations) one gets (\ref{d22})
\qed 

From \cite{Mar86} (theorem 1.5.1 p. 71)  we learn that for
each $\rho >0$, the Dirichlet spectrum of $-\partial_{xx} +V$ admits
an asymptotic expansion of the form
\begin{equation}  \label{expansion}
   \omega_{{j}} = j^{2} + c_{0}(V) + c_{1}(V)j^{-2} +\ldots + 
   c_{\rho}(V)j^{-2\rho} + C_\rho(V)j^{-2\rho-2}
\end{equation}   
   where $c_{0}(V) = \int_{0}^{2\pi}V(x)dx$ and the $c_{s}(V)$ are
   certain multilinear expressions in $V$ which depend smoothly on
   each of its Fourier coefficients (this is the only property we need).
   Moreover one has $|C_\rho(V)|<C(\rho)R$ for all potentials in
   $\V_R$ (here $C(\rho)$ denote a constant depending on $\rho$).

\begin{corollary}
\label{cs}
\begin{equation}
\label{cs1}
\left.\frac{\partial ^2 c_{s}}{\partial v_k\partial v_l}\right|
_{V=0} = \delta_{kl}\frac{1}{4} \frac{k^{2(s-1)}}{2^s}. 
\end{equation}\end{corollary}
\proof Just take the second derivative of equation (\ref{expansion}) and
compare with (\ref{d22}). \qed

It follows that the `constants' $c_s$ have an expansion at the origin
of the form
\begin{equation}
\label{exc}
c_s(v_1,v_2,...)=\sum_{k\geq1} A_{sk}v_k^2+w_s(v_1,v_2,...)
\end{equation}  
where the matrix $A_{sk}$ is given by the r.h.s of (\ref{cs1}) (without
the Kronecker symbol) and the functions $w_s$ have a zero of third
order at the origin. In particular there exists for each $k\geq 1$ a 
constant $C(k)$ such that
\begin{equation}\label{w}
|w_k(v_1,v_2,....)|\leq C(k) R^3\ .
\end{equation} 

\begin{remark}
\label{v.1111}
For any $\rho$, the matrix
$$
A:=(A_{sk})_{k=1,...,\rho}^{s=1,...,\rho}\ ,
$$
 is invertible and there exists a constant $C$ that depends
only on $\rho$ such that
\begin{equation}
\label{a-1}
\norma{A^{-1}}\leq C \ .
\end{equation}
(Indeed remark that the determinant of $A$ is a non vanishing 
Vandermonde determinant.)

\end{remark}

In the remainder part of this section we fix once for all $r\geq 1$. 
We begin with the following simple lemma:
\begin{lemma}
\label{v.111}
Fix $\rho \geq l \geq 1$ two integers. For any $J\geq 2$ and any
positive $\mu$ consider an arbitrary collection of $l$ indexes
\begin{equation}
\label{indiJ}
J\leq j_1<j_2<...<j_l\leq J^{\mu}\ ,
\end{equation}
and define the matrix
\begin{equation}
\label{v.101}
B=(B_{is})_{i=1,...,l}^{s=1,...,\rho}\ ,\quad
B_{is}:=\frac{1}{j_i^{2s}} \ .
\end{equation}
There exists an $l\times l$ submatrix $B_l$ which is invertible
and fulfills
\begin{equation}
\label{v.105}
\norma{B_l^{-1}}\leq C  J^{\beta(\mu,l)}
\end{equation}
with $\beta (\mu,l):=\frac 3 2 \mu l(l-1)$ and $C\equiv C(l)$ is
independent of $J$ and of the choice of the indexes.
\end{lemma}
\proof  Define the submatrix
$B_l=(B_{is})_{i=1,...,l}^{s=1,...,l}$, notice that all the
coefficients of the comatrix of $B$ are smaller than $l^2$ and
therefore
$$
\norma{B_l^{-1}}_{\infty} \leq l^2 (\mbox{det } B_l)^{-1}\ .
$$
On the other hand,
the determinant of $B_l$ is a Vandermonde determinant whose value
is given by
$$
\prod_{1\leq i<k\leq l}\left(\frac{1}{j_i^{2}}-\frac{1}{j_k^{2}}\right) 
\prod _{1\leq k\leq l} \frac{1}{j_k^{2}} \ .
$$ By the limitation (\ref{indiJ}) each term of the first product is
estimated from below by $J^{-3\mu}$. Since the number of pairs is
$l(l-2)/4$ one has the result.  \qed

\begin{lemma}
\label{v.113}
Given $J\geq 2$ and $\mu >1$ consider an arbitrary
collection of indexes
\begin{equation}
\label{indiJ1}
J\leq j_1<j_2<...<j_l\leq J^{\mu}\ , \ l\leq r
\end{equation}
Let $(a_{j_i})_{i=1,...,l}$ be a vector with components in $\Z$
fulfilling
\begin{equation}
\label{v.106}
a\not =0\ \ , \sum_{i}|a_{j_i}|\leq r\ ,\ \sum_{i}
j_i^2a_{j_i}=0 \ .
\end{equation}
For any positive $\Gamma$ and $\alpha>2\beta (\mu,l)$ define
\begin{equation}
\label{v.107}
\resto :=\left\{ V\in \V_R \ :\
\left|\sum_{i}\lambda_{j_i}a_{j_i}\right|\leq
\frac{\Gamma}{J^{\alpha}}      \right\} \ .
\end{equation}
For $R$ small enough, there exists two constants $C_1(r,\alpha)$ and $C_2(r,\alpha)$  such that,
provided 
\begin{equation}
\label{v.108}
J>\frac{C_1}{\Gamma^{1/2}}
\end{equation} 
one has
$$
|\resto|\leq C_2\frac{\Gamma^{1/2}}{J^{\frac{\alpha}{2}-\beta}} 
$$
where $\beta \equiv \beta (\mu,l)$ is defined in the preceding lemma.
\end{lemma}
\proof In order to fix ideas take $l=r$ the other case being
similar. Consider 
$$
\sum_{i}\lambda_{j_i}a_{j_i}=
\sum_{i=1}^{r}a_{\ji}\sum_{s=1}^{\rho}\frac{c_s}{\ji^{2s}} +
\sum_{i=1}^{r}a_{\ji}\frac{C_{\rho}(V)}{\ji^{2\rho+2}} 
$$ The second term at r.h.s. is bounded by $\frac{RrC(\rho)}{J^{2\rho+2}}$, 
thus if $R\leq 1$,
$\rho=\alpha/2$ and (\ref{v.108}) is satisfied, it can be
neglected and it remains to estimate the measure of the set of 
potentials such that
\begin{equation}
\label{v.109}
\left|\sum_{i=1}^{r}a_{\ji}\sum_{s=1}^{\rho}\frac{c_s}{\ji^{2s}}
\right| <\frac{\Gamma}{2J^{\alpha}}\ .
\end{equation}
We will show that there exists $\bar k\leq \rho $ such that the second
derivative of (\ref{v.109}) with respect to $v_{\bar k}$ is bounded
away from zero, and we will apply lemma \ref{v.112}. One has 
\begin{eqnarray}
\sum_{i=1}^{r}a_{\ji}\sum_{s=1}^{\rho}\frac{c_s}{\ji^{2s}} =
\sum_{i=1}^{r}a_{\ji}\sum_{s=1}^{\rho}\sum_{k=1}^{\rho}
\frac{A_{sk}v_k^2}{\ji^{2s}}\nonumber
\\ 
\label{v.100}
+ \sum_{i=1}^{r}a_{\ji}\sum_{s=1}^{\rho}\frac{w_s}{\ji^{2s}}+b
\end{eqnarray}
where 
$$
b:=\sum_{i=1}^{r}a_{\ji}\sum_{s=1}^{\rho}
\sum_{k>\rho}\frac{A_{sk}v_k^2}{\ji^{2s}} 
$$
is independent of $v_k$, $k=1,...,\rho$. Define $\tilde w_k$ by
$$
w_s=\sum_{k=1}^{\rho}A_{ks}\tilde w_k\ ,
$$
so that 
$$
\left|\tilde w_k\right|\leq CR^3
$$
with a constant depending only on $\rho$ (cf. \eqref{w} and (\ref{a-1})). So,
(\ref{v.100}) can be written as
\begin{equation}
\label{v.102}
\sum_{k=1}^{\rho}(v_k^2+\tilde w_k)\sum_{s=1}^{\rho} A_{ks}
\sum_{i=1}^{r} B_{ij}a_{\ji}+b=\sum_{k=1}^{\rho}(v_k^2+\tilde w_k)f_k+b
\end{equation}
where $B$ is given by (\ref{v.101}) and $f_k$ is defined as the
coefficient of the bracket at l.h.s. Clearly one has $f=ABa$. Now, the
image of $\R^r$ under $AB$ is an $r$ dimensional space (due to
remark \ref{v.1111} and lemma \ref{v.111}) and $AB$ is an isomorphism between such
$r$ dimensional spaces. By remark \ref{v.1111} and lemma \ref{v.111}
its inverse $(AB)^{-1}$ is bounded by
$$
\norma{(AB)^{-1}}\leq CJ^{\beta}\ ,
$$
(with $C$ independent of $J$), and therefore one has
$$
\norma{f}\geq \frac{1}{CJ^{\beta}}\norma a
$$
It follows that there exists $\bar k$ with $\left| f_{\bar
  k}\right|\geq \frac{1}{CJ^{\beta}}$. 

Consider now the second derivative of (\ref{v.102}) with respect to
$v_{\bar k}$, it is given by
\begin{equation}
\label{v.103}
f_{\bar k}\left(1+\frac{\partial^2\tilde w_k }{\partial v_{\bar k}^2}
\right) \ .
\end{equation}
Provided 
$$
\left| \frac{\partial^2\tilde w_k }{\partial v_{\bar k}^2}
\right|<\frac{1}{2} 
$$
which is a smallness assumption on $R$ (independent of $J$),
(\ref{v.103}) is larger than $\frac {1}{CJ^{\beta}}$. Applying lemma
\ref{v.112} one gets the thesis. \qed

In the next lemme, we relax the hypothesis \eqref{indiJ1} of the 
previous lemma.
\begin{lemma}
\label{diof}
There exists $R$ and
a positive constant $C$ with the following property:
for any positive $\Gamma$ there exists a  set $\Sc_1\subset \V_R$ such
that
\begin{itemize}
\item[i)] denote $\alpha:=4^r(r!)^2$, and let $J$ be an integer
fulfilling 
\begin{equation}
\label{j+}
J>\frac{C}{\Gamma^{1/2}}
\end{equation}
then for any
$V\in \Sc_1$ one has for $N\geq J$
\begin{equation}
\label{stimadiofricor1}
\left|\sum_{i=1}^{r}\lambda_{j_i}a_{j_i}\right|\geq
\frac{\Gamma}{N^{\alpha}} 
\end{equation}
for any choice of integers $J\leq j_{1},\ldots,j_{r}\leq N$ and of
relative integers $\{a_{j_i}\}_{i=1}^r$ fulfilling
$$
0\not=\sum_i |a_{j_i}|\leq r \ .
$$
\item[ii)] $\Sc_1$ has large  measure, namely
\begin{equation}
\label{misn}
{|\V_R-\Sc_1|}\leq C \Gamma^{1/2} 
\end{equation} 
\end{itemize}
\end{lemma}

\proof We prove the lemma by induction on $r$. More
precisely we prove that for any $l\leq r$ there exists a set $\N_l$
and two constants 
$\gamma_l, \alpha(l)$ such that 
\begin{equation}
\label{stimadiofricor}
\left|\sum_{i=1}^{l}\lambda_{j_i} a_{j_i}\right|\geq
\frac{\gamma_l}{J^{\alpha(l)}}
\end{equation}
for any $\V\in\N_l$ and moreover the measure estimate holds.  The
statement is true for $l=1$. Suppose it is true for $l-1$, we prove it
for $l$. Assume that all the integers $a_{j_1},...,a_{j_l}$ are
different from zero, otherwise the result follows by the inductive
assumption. Order the indexes in increasing order. First
remark that if $
\sum_{i=1}^{l}j_i^2a_{j_i}\not=0$ then
$\left|\sum_{i=1}^{l}j_i^2a_{j_i}\right|\geq 1 
$.
Hence, since the potential is small (for $R$ small enough) the l.h.s. of
(\ref{stimadiofricor}) is larger than 1/2 and the result holds in 
this particular case. Thus
we consider only the case where $\sum_{i=1}^{l}j_i^2a_{j_i}=0$. 

Now, since 
$$
\left|\lambda_{\ji}-\ji^2\right|\leq \frac{C}{\ji^2}\ ,
$$
if we assume that 
 $j_l^2>\frac{2rCJ^{\alpha(l-1)}}{\gamma_{l-1}}$, then we deduce from the
inductive assumption that
$$
\left|\sum_{i=1}^{l}\lambda_{j_i} a_{j_i}\right|\geq
\left|\sum_{i=1}^{l-1}
\lambda_{j_i} a_{j_i}\right|- \frac{rC}{j_l^2}\geq
\frac{\gamma_{l-1}}{2J^{\alpha(l-1)}}\ . 
$$
and thus, provided $\gamma_{l}\leq \frac {\gamma_{l-1}} 2$ and 
$\alpha_{l}\geq \alpha_{l-1}$, the lemma holds. So, 
assuming
\begin{equation}
\label{Jgrand}
J>\frac{C(r)}{\gamma_{l-1}^{1/2}}\ ,
\end{equation}
we have only to consider 
the case 
$
j_l< J^{\mu_{l}}
$
where $\mu_{l}=[\alpha(l-1)+2]/2$.

We now apply  lemma \ref{v.113}. The number
of possible choices for the integers $a_{j_1},\ldots,a_{j_{l}}$ is 
smaller than $J^{l\mu_{l}}$.Therefore,  the measure of the set of 
potentials that we have to exclude at the
$l$-th step can be estimated by
$$
\frac{C\gamma_l^{1/2}J^{l\mu_{l}}}{J^{\frac{\alpha(l)}2-\beta(\mu_{l},l )}}
$$
Thus in order to obtain the result it suffices to choose $\alpha_{l}$ 
in such way that
$$
\alpha(l)>\alpha(l-1)(3l^{2}+l)
$$ which amounts to the requirement of a sufficiently fast growth of
$\alpha(l)$. The choice $\alpha(l)=4^l(l!)^2$ fulfills this
requirement. On the other hand, choosing $\gamma_{r}= \Gamma$,
$\gamma_l=\gamma_{l-1}/2$ for $l=2,\ldots,r$ and assuming \eqref{j+}
with a suitable constant $C$, \eqref{Jgrand} is satisfied for each
$l=2,\ldots,r$. This gives the statement of the lemma with a suitable
constant in (\ref{misn}) and with $J$ in place of $N$ in
(\ref{stimadiofricor1}). As $J\leq N$, the lemma is proved.  \qed

We now take into account the small indexes ($j<J$):

\begin{lemma}
\label{v.212}
There exists $R>0$ with the following properties: for any
$\gamma>0$ and $ J>0$ fulfilling 
\begin{equation}
\label{Jgamma}
\frac{\gamma e^{2\sigma J}}{R}<1
\end{equation} 
there exists a set $\Sc_J\subset \V_{R_*} $ such that
\begin{itemize}
\item[i)] $\forall V\in\Sc_J$ one has
\begin{equation}
\label{v.300}
\left|\sum_{j=1}^{J}\lambda_jk_j+ \sum_{j=J+1}^{N}\lambda_ja_j\right|
\geq \frac{\gamma}{N^{r+2}}\ ,\ \forall N\geq J 
\end{equation}
and for all $k\in\Z^J$, $a\in Z^{N-J}$ fulfilling $|k|+|a|\leq r$,
$k\not=0$. 
\item[ii)] $\left|\V_{R_*}-\Sc_J\right|\leq \displaystyle{\frac{\gamma
  e^{2\sigma J}}{R}}$.
\end{itemize}
\end{lemma}
\proof To start with fix $k$ and $a$. We estimate the measure of the
set such that (\ref{v.300}) is violated. Let $i\leq J$ such
that $k_i\not=0$. Take the derivative of the argument of the modulus
in (\ref{v.300}) with respect to $v_{2i}$. By (\ref{d.21}) this is
given by $\frac{-1}{2}k_i+\resto$ with $|\resto|<CR<1/4 $ provided $R$ is
small enough. It has to be remarked that, here, $C$ does not depend on
$J,a,\gamma$ but only on $r$. Thus the measure of the set such that (\ref{v.300}) is
violated for the given choice of $(k,a)$ is estimated by
$\frac{\gamma}{4N^{r+2}}\frac{ e^{2\sigma J}}{R}$. Summing over all
the possible choices of $k$ and $a$ (namely $N^r$)
and summing over $N$,
one gets the thesis. \qed 

Combining the last two lemmas one gets

\begin{theorem}
\label{a.nls}
Fix $r\geq 1$ and $\alpha = 4^r (r!)^2$. There exists $R$ with the following property:
$\forall\gamma>0$ there exists a set $\Sc\subset \V_{R}$ such that 
\begin{itemize}
\item[i)] for any $ V\in\Sc$ one has for $N\geq 1$
\begin{equation}
\label{v.307}
\left|\sum_{j=1}^{N}\lambda_jk_j\right|
\geq \frac{\gamma}{N^{\alpha}}\ ,
\end{equation}
for all $k\in\Z^N$ with $0\not= |k|\leq r$. 
\item[ii)] $\left|\V_{R}-\Sc\right|\leq
  \displaystyle{\left[C\ln\left(\frac{R^2
  }{\gamma}\right)\right]^{-1/2} }$.
\end{itemize}
\end{theorem} 
\proof With a slight change of notation write (\ref{v.307}) in the
form 
\begin{equation}
\label{v.201}
\left|\sum_{j=1}^{J}\lambda_jk_j+ \sum_{j=J+1}^{N}\lambda_ja_j\right|
\geq \frac{\gamma}{N^{\alpha}}
\end{equation}
with a still undetermined $J$.  To begin with consider the case $k=0$
and apply lemma \ref{diof} with a still undetermined $\Gamma$. In
order to fulfill (\ref{j+}) we choose now $J:= \frac{2C}{\Gamma^{1/2}}$. It
follows that (\ref{v.201}) holds for any $\gamma\leq \Gamma $ 
provided $V$ belongs to a set
$\Sc_1$ whose measure is estimated by $C\Gamma^{1/2}$. If $k\not=0$ 
then assuming
$$
\frac{\gamma e^{2\sigma J}}{R}= \frac{\gamma e^{\frac{2\sigma
      C}{\Gamma^{1/2}}}}{R}<1 
$$
with $R\equiv R(r)$ chosen as in lemma \ref{v.212}, 
one can apply lemma \ref{v.212} 
and (\ref{v.201}) is still satisfied. 

Now we choose  $\Gamma$ in such a
way that 
$$
\frac{\gamma e^{\frac{2\sigma
      C}{\Gamma^{1/2}}}}{R}=\gamma^{1/2}
$$
namely 
\begin{equation}
\label{v.202}
\Gamma^{1/2}=\frac{4\sigma C}{\ln(\frac{R^2}{\gamma})}\ .
\end{equation}
It follows that $\gamma<\Gamma$ and therefore (\ref{v.201}) is true in
$\Sc_J\cap \Sc_1$, whose complement is estimated by twice the
r.h.s. of (\ref{misn}) with $\Gamma$ given by (\ref{v.202}).\qed 

Finally, to deduce theorem \ref{NLSdir} from theorem \ref{a.nls}
just adapt the proof of lemma \ref{v.004}. To this end, instead of
(\ref{v.007}) use that, due to the asymptotic of the
frequencies, $\omega_j-\omega_l>l/2$. 

\subsection{Frequencies of coupled NLS} \label{couple}

Denote $\omega^{(N)}:=(\omega_{-N},...,\omega_N)$, then, by repeating
the procedure of the previous subsection, one gets the following
\begin{theorem}
\label{a1.nls}
Fix $r\geq 1$ and $\alpha_1 = 4^r (r!)^2$. There exists $R$ with the
following property: $\forall\gamma_1>0$ there exists a set $\Sc\subset
\V_{R}\times\V_R$ such that
\begin{itemize}
\item[i)] for any $ (V_1,V_2)\in\Sc$ one has for $N\geq 1$
\begin{equation}
\label{v.1307}
\left|\omega^{(N)}\cdot k\right|
\geq \frac{\gamma_1}{N^{\alpha_1}}\ ,
\end{equation}
for all $k\in\Z^{2N}$ with $0\not= |k|\leq r$. 
\item[ii)] $\left|\V_{R}-\Sc\right|\leq
  \displaystyle{\left[C\ln\left(\frac{R^2
  }{\gamma_1}\right)\right]^{-1/2} }$.
\end{itemize}
\end{theorem} 

\noindent{\it Proof of theorem \ref{cou.9}}. As in the proof of lemma
\ref{v.004} we study the inequality
\begin{equation}
\label{v.1005}
\left|\omega^{(N)}\cdot
k+\epsilon_1\omega_j+\epsilon_2\omega_l\right|\geq \frac{\gamma}{N^{\alpha}}
\end{equation}
with $k\in\Z^{2N}$, $\epsilon_i=0,\pm1$, $|j|\geq |l|>N$.  The cases
$\epsilon_1=\epsilon_2=0$ and $\epsilon_1\epsilon_2=1$ can be treated
exactly as in the proof of lemma \ref{v.004}; the same is true when
$\epsilon_1\epsilon_2=-1$ and $j$ and $l$ have the same sign. In such
cases one gets that (\ref{v.1005}) holds with $\alpha=2\alpha_1$ in a
set whose complement is estimated by item ii of the theorem with a
different constant $C$. Consider now the case
$\epsilon_1\epsilon_2=-1$, $j>0$, $l<0$. If $j\not=-l$ then the same
argument as before applies. Take now $l=-j$ and, to fix ideas, take
$\epsilon_1=1$ and $\epsilon_2=-1$, and write $\omega_{\pm j}=\pm
j^2+\nu_{\pm j}$ with $|\nu_{\pm j}|\leq Cj^{-2}$. Thus the argument
of the modulus at l.h.s. of eq.(\ref{v.1005}) reduces to
$$
\omega^{(N)}\cdot
k+\nu_j-\nu_{-j}\ .
$$
If $j^2\geq 2CN^{\alpha_1}/\gamma_1$ then the $\nu$ terms can be
neglected, and, provided $k\not=0$ the estimate
(\ref{v.1005}) follows. We have now to consider the case $j^2<
2CN^{\alpha_1}/\gamma_1$. To get the estimate (\ref{v.1005}) simply
apply (the procedure leading to) theorem \ref{a1.nls} with $N$
substituted by $2CN^{\alpha_1}/\gamma_1$. This leads to (\ref{v.1005})
with $\alpha=\alpha_1^2/2$ and $\gamma=C\gamma_1^{\alpha_1/2
  +1}$. \qed

\subsection{Frequencies of NLS in higher dimension}

Denote $\omega^{(N)}:=(\omega_j)_{|j|\leq N}$.

\begin{lemma}
\label{lem.res.d}
Fix $r\geq 1$ and $\gamma >0$ small enough.  There exist positive
constants $C\equiv C_{r}$, $\beta \equiv \beta (r,\gamma)$ and a set
$\Sc_{\gamma} \subset \V$  with $\va{\V-\Sc_{\gamma}}\to 0$ when 
$\gamma \to 0$
such
that, if $V\in \Sc_{\gamma}$ then for any $N\geq 1$ and any $n\in \Z$
\begin{eqnarray}
\label{nr.d2}
|\omega^{(N)}\cdot
k +n|\geq  \frac{\gamma}{N^\beta}\ , 
\end{eqnarray}
for any $k\in\Z^{\Z^d}$ with $0<|k|\leq r$.
\end{lemma}

\proof First of all recall that for $V\in \V$, $V(x)=\sum_{k\in
\Z^d} {v_{k}}\ e^{\im k\cdot x}$, the frequencies
are given by
$$
\omega_{k} = \vak^{2} +{v_{k}} \ .
$$
Then notice that, given $(a_{1},\ldots,a_{r}) \neq 0$ in $\Z^r$, 
$M>0$ and 
$c\in \R$ the Lebesgue measure of 
$$
\{x\in[-M,M]^r \mid \va{\sum_{i=1}^r a_{i}x_{i} +c}<\delta \}
$$
is smaller than $(2M)^{r-1} \delta $.  Hence given $k\in \Z^{\Z^d}$ of
length less than $r$ and $n\in \Z$ the Lebesgue measure of
$$
\mathcal X_k:=\left\{ x\in[-1/2,1/2]^{\Z^d}\ :\ \left|\sum_{|j|\leq N}
k_j(|j|^2+x_j)+n \right|<\frac{\gamma}{N^{\beta}}  \right\}
$$
is smaller than $\gamma/N^\beta$. Therefore its probability measure is
estimated by $(1+N)^{m(r-1)}\gamma/RN^\beta$. To conclude the proof we have
to sum over all the $k$'s and the $b$'s. To count the cardinality of
the set of the $k$'s and the $n$'s to be considered remark that if
$|\omega^{(N)}\cdot k -n|\geq \delta$ with $\delta <1$ then
$\va{b}\leq 1 + \va{\omega^{(N)}\cdot k} \leq 1 + (1+N^{2})r$. So that  
to ensure 
\eqref{nr.d2} for all possible choices of $k$, $n$ and $N$, one has 
to remove from $\V$ a set of measure
$$
\sum_{N\geq 1} \gamma \frac{1}{N^\beta}(1+N)^{m(r-1)}(2N)^{dr}
(1 + (1+N^{2})r) \ .
$$
Choosing $\beta := r(d+m)+5$, the last series converges and the lemma 
is proved. \endproof 

To deduce theorem \ref{NLSdir} from lemma \ref{lem.res.d}
one has just to adapt the proof of lemma \ref{v.004}.

\appendix
\section{Technical Lemmas}\label{A}
\label{tech}

\subsection{Proof of lemma \ref{Ngrande}.} 
Introduce the projector
$\bar\Pi$ on the modes with index smaller than $N$ and the
projector $\hat \Pi$ on the modes with larger index. Expand $f$
in Taylor series (in all the variables), namely write  
$$
f=\sum_{j\leq r_*+2}f_{j}
$$
with $f_{j}$ homogeneous of degree $j$. 
Consider the vector field of $f_{j}$ and decompose it into the component on
$\bar z $ and the component on $\hat z$. One has:
\begin{eqnarray}
\label{s.N1}
\bar \Pi X_{f_{j}}= J_{\bar z} \nabla_{\bar z }f_{j}\ ,
\\
\label{s.N2}
\hat \Pi X_{f_{j}}= J_{\hat z}\nabla_{\hat z}f_{j}\ ,
\end{eqnarray}  
where we denoted by $J_{\bar z}$ and $J_{\hat z}$ the two components
of the Poisson tensor. From(\ref{s.N1},\ref{s.N2}) one immediately
realises that $\bar \Pi X_{f_{j}}$ has a zero of order three as a
function of $\hat z$ and $\hat \Pi X_{f_{j}}$ has a zero of order two
as a function of $\hat z$. Consider $\hat \Pi X_{f_{j}}$, write
$z=\bar z +\hat z$, one has
\begin{eqnarray}
\label{s.N3}
\hat \Pi X_{f_{j}}(\bar z +\hat z)= \hat \Pi \widetilde{X}_{f_{j}}
({\underbrace{\bar z +\hat z,...,\bar z +\hat z}_{(j-1)\text{--times}} })
\\
\nonumber
=\sum_{l=2}^{j-1}\newton{j-1}l \hat \Pi \widetilde{X}_{f_{j}}
({\underbrace{\hat z,...,\hat z}_{l\text{--times}} }\hskip5pt , 
{\underbrace{\bar z ,...,\bar z }_{(j-l-1)\text{--times}} }
)
\end{eqnarray}
where the sum starts from $2$ since $\hat \Pi X_{f_{j}}$ has a
zero of order two as a function of $\hat z$. We estimate now a single
term of the sum. By the tame property we have
\begin{eqnarray}
\label{s.N4}
\norma{\hat \Pi \widetilde{X}_{f_{j}} ({\underbrace{\hat z,...,\hat
z}_{l\text{--times}} }\hskip5pt , {\underbrace{\bar z
,...,\bar z }_{(j-l-1)\text{--times}} } ) }_s 
\\ 
\nonumber 
\leq
|X_{f_{j}}|_s^T \frac{1}{j-1}\left[\sum_{i=1}^{l}\norma{\hat z}
^{l-1}_{1} \norma{\hat z}_s\norma{\bar z }_{1}^{j-1-l}
+\sum_{i=l+1}^{j-1}\norma{\hat z} ^{l}_{1} \norma{\bar z
}_s\norma{\bar z }_{1}^{j-2-l} \right]
\end{eqnarray}
Using the inequalities
\begin{eqnarray*}
\norma{\hat z}_{1}&\leq&\frac{\norma{\hat z}_s}{N^{s-1}} \ ,
\\
\norma{\hat z }_s&\leq& \norma{z }_s\ ,\quad \norma{\bar z  }_s\leq
\norma{z }_s 
\\
\norma{z }_{1}&\leq& \norma{z}_s
\end{eqnarray*}
which immediately follow from the definition of the norms one can
estimate the quantity (\ref{s.N4}) by
$$
\frac{1}{N^{(s-1)(l-1)}}\norma{z}_{s}^{j-1}
$$
  Inserting into (\ref{s.N3}) one gets for $z\in B_s(R)$
$$
\norma{\hat \Pi{X}_{f_{j}}(z)}_s\leq 2^{j-1}
\frac{R^{j-1}}{N^{s-1}}|X_{f_{j}}|_s^T \ .
$$
 A similar estimate with $N^{2(s-1)}$ instead of
$N^{s-1}$ holds for $\bar \Pi {X}_{f_{j}}(z)$. Therefore
$$
\norma{{X}_{f_{j}}(z)}_s\leq 2^{j}
\frac{R^{j-1}}{N^{s-1}}|X_{f_{j}}|_s^T 
$$
and summing over $j$ one gets the thesis. \qed

\subsection{Proof of lemma \ref{poisson}} One has
$$
X_{\{f,g\}}=[X_f,X_g]=dX_f X_g-dX_gX_f
$$ Denote $X:=X_f$, $Y:= X_g$. Write 
$$
X(z)=\sum_{k,l_1,...l_{n}}X_k^{l_1,...,l_{n}}\be_kz_{l_1}...z_{l_{n}}
$$
and similarly for $Y$, then one has
\begin{eqnarray*}
\mod{[X,Y]}(z) \\ = \sum_{{k,l_1,...l_{n}\atop j_1,...j_{m}}}\left|
nX_k^{l_1,...,l_{n}} Y_{l_n}^{j_1,...,j_{m}}- m
X_{l_n}^{l_1,...,l_{n-1},j_m} Y_{k}^{j_1,...,j_{m-1},l_n} \right|
\be_kz_{l_1}...z_{l_{n-1}}z_{j_1}...z_{j_{m}}
\end{eqnarray*}
After symmetrization with respect to the indexes
$l_1,...,l_{n-1},j_1,...,j_{m}$, 
the quantities in the above modulus are the components of a
multilinear symmetric form $\widetilde{\mod{[X,Y]}}$. Consider also the
multilinear form with components obtained by symmetrizing the
quantities 
\begin{equation}
\label{1.2}
\sum_{l_n}\left(\left|n X_k^{l_1,...,l_n}
Y_{l_n}^{j_1,...,j_{m}}\right|+\left| m X_{l_n}^{l_1,...,l_{n-1},j_m}
Y_{k}^{j_1,...,j_{m-1},l_n} \right|\right)
\end{equation}
namely the form $\widetilde{d\mod X\mod Y}+\widetilde{d\mod Y\mod X}$.
Let $w=(z^{(1)},...,z^{(n+m-1)})$ be a multivector with all the
vectors $z^{(j)}$ chosen in the positive octant (see remark
\ref{sup}). Then the value of $\widetilde{\mod{[X,Y]}}$ on $w$ is
bounded by the value of $\widetilde{d\mod X\mod Y}+\widetilde{d\mod
Y\mod X}$ on the same multivector and   finally it suffices to estimate 
$\widetilde{d\mod X\mod Y}$.

The value of $\widetilde{d\mod X\mod Y}$ on the multivector
$w$ is given by 
\begin{equation}
\label{symme}
n\sum_{\sigma} \frac{1}{(n+m-1)!}\widetilde{ \mod
X}\left(z^{(\sigma(1))},..., z^{(\sigma(n-1))},\widetilde{\mod
Y}(z^{(\sigma(n))},...,z^{(\sigma(n+m-1))} ) \right)
\end{equation}
where $\sigma$ are here all the permutations of the first $n+m-1$
integers. 

  Consider one of the terms in the last sum, say the one corresponding 
to
the identical
permutation. One has
\begin{equation}
\label{permu1}
\norma{\widetilde{ \mod
X}\left(z^{(1)},..., z^{(n-1)},\widetilde{\mod
Y}(z^{(n)},...,z^{(n+m-1)} ) \right)}_s\leq |\mod X|_s^T|\mod Y|_s^T
\sum_{l=1}^{n+m-1} c_ly_l\ , 
\end{equation}
where
$$ y_l:=\norma{z^{(1)}}_{1}...\norma{z^{(l-1)}}_{1}
\norma{z^{(l)}}_{s}
\norma{z^{(l+1)}}_{1}...\norma{z^{(n+m-1)}}_{1}
$$
and
\begin{eqnarray*}
c_l&:=&\frac{1}{n}\ ,\qquad\text{if}\quad l=1,...,n-1
\\
c_l&:=&\frac{1}{mn}\ ,\quad\text{if}\quad l=n,...,n+m-1\ .
\end{eqnarray*}
Now the important property is that $\sum_lc_l=1$. Consider now the
norm of the sum in (\ref{symme}), it is clear that it is again
estimated by an expression of the form of the r.h.s. of (\ref{permu1})
with suitable constants $c'_l$. To compute the constants $c'_l$ remark
that, due to the symmetry of the expression all the coefficients must
be equals. Moreover, the symmetrization does not change the property
that the sum\ of the coefficients is 1, and therefore one has
$c'_l=1/(n+m-1)$ for all $l$'s. So, in conclusion one has 
$$
\norma{\widetilde{d\mod X\mod Y}(w)}_s\leq n |\mod
X|_s^T |\mod
Y|_s^T \norma{w}_{s,1}\ ,
$$
from which one gets
the thesis.\qed

\section{On the verification of the 
tame modulus property}
\label{ApT}

It is clear that the linear combination of polynomials
with tame modulus still has tame modulus. The tame modulus property 
is also stable by composition:

\begin{lemma}
\label{compo}
Let $X:\Ph_s\to\Ph_s$ and $Y:\Ph_s\to\Ph_s$ be two polynomial vector
fields with $s$-tame modulus, then also their composition $X\circ Y$
has $s$-tame modulus. 
\end{lemma}
\proof
Just remark that for any multivector $w$ with all components in the
positive octant one has 
$$
\norma{\widetilde{\mod{X\circ Y}}(w)}_s\leq \norma{\widetilde{\mod
    X\circ\mod Y}(w)}_s 
$$
and that the composition of tame maps is
still a tame map, so that 
$$
\norma{\widetilde{\mod{X\circ Y}}(w)}_s\leq C_s\norma{w}_{1,s} 
$$
with a suitable $C_s$.\qed

Now the idea is that if one is able to show that some elementary maps
have tame modulus(think of examples \ref{ex.2}, \ref{ex.21}),
then the same is true for the maps obtained by composing them.

We have already seen (cf example \ref{linear}) that bounded linear
maps are tame.  Proposition \ref{localizzato} below gives a simple
condition to ensure that a linear map has tame modulus for all
$s\geq 1$. Such a condition will be very useful in studying the behaviour
of the $T_M$ property under change of basis in $\Ph_s$. In the
following proposition, given a linear operator $A$, we will consider
its matrix defined by $Az=\sum_{kl}A_{kl}z_l\be_k$. We
will simply write $A=(A_{kl})$.

\begin{proposition}
\label{localizzato}
Let $a$ be an injective and surjective map from $\Zb$ to $\Zb$
with the property that there exists $C>0$ such that
\begin{equation}
\label{l}
\frac {|l|}C\leq |a(l)|\leq C|l|
\end{equation}
and let 
$A=(A_{kl})$ be a linear operator. If for any $n$
there exists a constant $C_n$ such that
\begin{equation}
\label{locai}
|A_{kl}|\leq \frac{C_n}{(1+\left|k- a(l)\right|)^{n}}
\end{equation} 
then $A$ has tame modulus.
\end{proposition}
\proof First we get rid of the map $a$ by reordering the basis in the
space $\Ph_s$ constituting the domain of $A$. Namely we take as $l$-th
element of the basis $\be_{a(l)}$. This has the consequence of
changing the norm in the domain; due to (\ref{l}) the new norm is
equivalent to the old one.  So we consider only the case where $a$ is
the identity.

  To prove that $A$ has tame modulus it suffices to prove that
$|A_{kl}|$ are the matrix elements of an operator which is bounded as
an operator from $\Ph_s$ into itself. This is clearly equivalent to
the fact that $D_{kl}:=|A_{kl}| |k|^{s} |l|^{-s}$ is bounded from
$\Ph_0$ into itself. Remark that, since
$$
\frac{|k|}{(1+\left|k-l\right|)|l|}\leq 1\ ,
$$ in view of \eqref{locai}, for any $n$ there still exists a
constant $C'_n$ such that
\begin{equation}
\label{dkl}
D_{kl}\leq\frac{C'_n}{(1+\left|k-l\right|)^n}\ .
\end{equation}
Thus, using Schwartz inequality, one has
\begin{eqnarray*}
\norma{Dz}_0^2=\sum_k\left(\sum_l D_{kl}z_l \right)^2\leq 
\sum_k\left[\left(\sum_l D_{kl} \right)\left(\sum_l D_{kl}z_l^2
  \right) \right]
\\
= K_n \sum_l |z_l|^2\left(\sum_k D_{kl}\right) =K_n^2
\norma z^2_0 
\end{eqnarray*}
where we used the inequality (\ref{dkl}) and we defined 
$$
K_n:=\sum_k \frac{C'_n}{(1+\left|k-l\right|)^n}\ .
$$
  \qed

In the particular case of linear {\it canonical} maps one has 

\begin{proposition}
\label{canonical}
Let $A^{(1)}...,A^{(m)}$ be linear operators fulfilling
\begin{equation}
\label{locai1}
|A^{(j)}_{kl}|\leq \frac{C_n}{(1+\left|k- a_j(l)\right|)^{n}}
\end{equation} 
where the maps $a_j$ have the property   \eqref{l}.  Let
$A:=A^{(1)}+...+A^{(m)}$ be a canonical map (which by the
proposition \ref{localizzato} has tame modulus) then also $A^{-1}$ has
tame modulus.
\end{proposition}
\proof Denote by $J$ the Poisson tensor, namely the operator $z\equiv
(p_k,q_k)\mapsto J(p_k,q_k):=(-q_k,p_k)$, which clearly has tame
modulus. Then, by canonicity one has
$$
A^{-1}=-JA^TJ=(-JA^{(1)}\null^TJ-...-JA^{(m)}\null^TJ)\ .
$$ 
 But $A^{(j)}\null^T$ still fulfills (\ref{locai1}) with the roles
of $k$ and $l$ exchanged, therefore it has tame modulus. It follows
that the above expression has tame modulus.\qed

  Finally concerning changes of coordinates one has the following

\begin{corollary}
\label{cambiobase}
Let $z=Aw$ be a linear canonical
transformation with the same structure as in proposition
\ref{canonical}. 
If $f\in T_M^s$ is a polynomial function with $s$-tame modulus, then
also the transformed function $f\circ A$ has $s$-tame modulus.   
\end{corollary}
\proof Just remark that $X_{f\circ A}= A^{-1}X_f\circ A$ which is the
composition of maps with $s$-tame modulus.\qed

\medskip

\noindent {\bf Proof of theorem \ref{prop:tame}.} First use the
isomorphism (\ref{isom}) to identify $\B_s$ with $\Ph_s$. By 
hypothesis $P$
has tame modulus in this basis. Then one has to pass to the basis of
the normal modes, namely to introduce the coordinates relative to the
basis $\varphi_j$. The matrix realizing the change of coordinates is
canonical and has the form
$$
\left( \begin{matrix}
    B & 0  \\
    0 & B
\end{matrix} \right)
$$
where the matrix elements of $A$ are given by  $B_{k,l}= \varphi_{l}^k$.
Therefore if (S1) is satisfied,
$$
\va{B_{k,l}} \leq \frac{C_n}{(1+|k-l|)^n} +\frac{C_n}{(1+|k+ l|)^n}\ .
$$
Thus, as a consequence of corollary \ref{localizzato}, one gets the 
thesis.
\qed


\bibliography{kamnew}
\bibliographystyle{amsalpha}

\bigskip
{
{\bf Dario Bambusi}

{\it Dipartimento di  Matematica

Via Saldini 50, 20133 Milano, Italy

\smallskip
E-mail: {\tt dario.bambusi@unimi.it}
\smallskip

 web-address}: {\tt http://users.mat.unimi.it/users/bambusi/ }
}

\smallskip

\smallskip

{\bf B\'enoit Gr\'ebert}

{\it Laboratoire de Mathematique Jean Leray,

Universit\'e de Nantes,

2, rue de la Houssini\`ere,

44322 Nantes Cedex 3, France

\smallskip
E-mail: {\tt grebert@math.univ-nantes.fr}
}

\end{document}